\documentclass[preprint,onecolumn,usenatbib,useAMS]{mn2e}

\usepackage{natbib}
\usepackage{amsbsy}
\usepackage{amssymb}
\usepackage{amsmath}
\usepackage{graphicx}
\usepackage[justification=centering]{caption}
\usepackage{mathrsfs}
\usepackage[caption = false]{subfig}
\usepackage{array}

\usepackage{xcolor}
\usepackage{hyperref}

\hypersetup{
  colorlinks,
  citecolor=blue,
  linkcolor=blue,
  urlcolor=blue}

\usepackage [english]{babel}
\usepackage [autostyle, english = american]{csquotes}
\usepackage{natbib}
\MakeOuterQuote{"}

\date{}
\def\be{\begin{equation}}
\def\ee{\end{equation}}
\def\v{\cal{V}}
\def\u{{\mathbf{U}}}
\def\S{\mathcal {S}}
\def\A{\mathcal {A}}

\def\N{\mathcal {N}}
\def \t{{\boldmath{\vec \theta}}}
\def\k{\mathbf{k}}
\def\kperp{\mathbf{k}_{\perp}}  
\def\kpar{k_{\parallel}}
\def \nh{H~{\sc i~}}

\begin{document}

\title[]{Post-reionization \nh 21-cm signal: A probe of negative cosmological constant}

\author[Dash,Guha Sarkar, Sen]{ Chandrachud B. V. Dash$^1$ \& Tapomoy Guha Sarkar$^{1}$ \& Anjan A. Sen$^{2}$\\ \\
 $^1$Department of Physics, Birla Institute of Technology and Science, Pilani Campus, Pilani, Jhunjhunu 333031, Rajasthan, India 
 $2$ Centre for Theoretical Physics, Jamia Millia Islamia, Delhi, 110025, India
 }
 
\author[Dash, Guha Sarkar, Sen ]{ Chandrachud B.V. Dash,$^{1}$\thanks{E-mail: cb.vaswar@gmail.com} Tapomoy Guha Sarkar,$^{1}$\thanks{E-mail: tapomoy1@gmail.com} Anjan A. Sen, $^{2}$\thanks{E-mail:aasen@jmi.ac.in}
\\
$^{1}$Birla Institute of Technology \& Science, Pilani, Jhunjhunu 333031, Rajasthan India\\
$^{2}$Centre for Theoretical Physics, Jamia Millia Islamia, Delhi, 110025, India\\ }

\pubyear{2023}

\maketitle

\begin{abstract}
In this study, we investigate  a  cosmological model involving a negative cosmological constant (AdS vacua in the dark energy sector). We  consider a  quintessence field on top of a negative cosmological constant and study its impact on cosmological evolution and structure formation. We use the power spectrum of the redshifted HI 21 cm  brightness temperature maps  from the post-reionization epoch as a cosmological probe. The signature of baryon acoustic oscillations (BAO) on the multipoles of the power spectrum is used to  extract measurements of the angular diameter distance $D_A(z)$ and the Hubble parameter $H(z)$. The projected errors on these are then  subsequently employed to forecast the constraints on the model parameters ($H_0, \Omega_{m}, \Omega_\Lambda, w_0, w_a$) using  Markov Chain Monte Carlo techniques. We find that a  negative cosmological constant with a phantom dark energy equation of state (EoS) and a higher value of $H_0$  is viable from BAO distance measurements data derived from galaxy samples. We also find that BAO imprints on the 21cm power spectrum 
obtained from  a futuristic SKA-mid like experiment  yield a $1-\sigma$ error on a negative cosmological constant and  the quintessence dark energy EoS parameters to be $\Omega_\Lambda=-1.030^{0.589}_{-1.712} $ and   $w_0=-1.023^{0.043}_{-0.060}$, $w_a=-0.141^{0.478}_{-0.409}$ 
respectively.

\end{abstract}

\section{Introduction}
One of the most significant discoveries of the twenty-first century was the fact that the expansion of the Universe is accelerated \citep{amendola_tsujikawa_2010}. Several independent observations confirm the counter-intuitive phenomenon of dark energy \citep{riess1998observational, perlmutter2003supernovae, McDonald_2007, scranton2003physical, Eisenstein_2005}. Observations indicate that about $\sim 64\%$ of the universe's total energy budget is made up of dark energy, which has a large negative pressure and acts as a repulsive force against gravity \citep{Padmanabhan_2003, Ratra-Peebles_1988}. In the last few decades, cosmological observations have attained an unprecedented level of precision. The $\Lambda$CDM model \cite{Carroll_2001, Ratra-Peebles_1988, Bulletal_2016} provides a good description towards explaining most properties of a wide range of astrophysical and cosmological data, including distance measurements at high redshifts \citep{riess1998observational, perlmutter2003supernovae, padmanabhan2003theoretician}, the cosmic microwave background (CMB) anisotropies power spectrum \citep{spergel2007three}, the statistical properties of large scale structures of the Universe \citep{bull2016extending} and the observed abundances of different types of light nuclei \citep{schramm1998big, steigman2007primordial, cyburt2016big}. All these observations point towards an accelerated expansion history of the Universe.

Despite the overwhelming success of the $\Lambda$CDM model as a standard model explaining these diverse observations, it still leaves significant uncertainties and is plagued with difficulties  \citep{weinberg1989cosmological, burgess2015cosmological, PhysRevLett.82.896, copeland2006dynamics, di2021realm, Buchert_2016, abdalla2022cosmology, Anchordoqui_2021, schoneberg_2022}. This is motivated by a wide range of observational results which has been in tension with the model.
Some of the issues faced by $\Lambda$CDM cosmological model other than the theoretical issues like the fine-tuning problem \citep{weinberg1989cosmological} etc, are posed by observational anomalies. Some of these anomalies at  a $ > 2-3\sigma$ level are the Hubble tension \citep{di2021combined, riess2020expansion, saridakis2021modified,dainotti2021hubble,bargiacchi2023tensions}/ growth tension \citep{abbott2018dark, Basilakos_2017, joudaki2018kids}
CMBR anomalies \citep{akrami2020planck, schwarz2016cmb}, BAO discrepancy \citep{addison2018elucidating, cuceu2019baryon, evslin2017isolating} and many others \citep{perivolaropoulos2022challenges}.

A positive cosmological constant is sometimes interpreted as a
scalar field at the positive minimum of its potential by moving the term $\Lambda g_{\mu \nu}$ to the right-hand side of the Einstein's  equation to include it in the energy momentum tensor  $T_{\mu \nu}$.  A  Quintessence \citep{PhysRevLett.81.3067, brax1999quintessence, Caldwell_2005, nomura2000quintessence} scalar field, on the contrary, slowly rolls towards the  minimum in the positive part of the potential giving rise to a dynamical dark energy with a time dependent equation of state 
$w(a) = P_{DE}/ \rho_{DE}$.
Several reports of the Hubble tension \citep{di2016reconciling, di2020nonminimal, vagnozzi2020new, alestas2020h, anchordoqui2020h, banerjee2021hubble, di2021realm, schoneberg_2022} has led to the proposal of a wide range of dark energy models.
There are certain proposed quintessence models with an  AdS vacuum  \citep{dutta2020beyond, calderon2021negative, akarsu2020graduated, visinelli2019revisiting,ye2020hubble,yin2022small} which do not rule out the possibility of a negative $\Lambda$.
We have considered Quintessence models, with a  non zero vacuum, which can be effectively seen as as a rolling scalar field $\phi$  on top of a cosmological constant $\Lambda \neq 0$. The combination $\rho_{_{DE}} = \rho_{\phi} + \Lambda $ satisfying the energy condition $\rho_{DE} >0$ drives an accelerated expansion \citep{sen2023cosmological}. 

We consider the post-reionization HI 21 cm brightness temperature maps as a tracer of the underlying dark matter distribution and thereby a viable probe of structure formation.
The intensity mapping  \citep{Bull_2015} of the post-reionization \nh 21 cm signal \citep{bharad04} is a promising  observational tool to 
measure cosmological evolution and structure formation tomographically \citep{ Mao_2008, zaldaloeb, bharad04}. The 21-cm power spectrum is expected to be a storehouse of cosmological information about the nature of  dark energy \citep{param1, param2, param3, param4, Sarkar_2015, hussain2016prospects, Dash_2021, dash2022probing},  and several radio interferometers like the 
 SKA\footnote{https://www.skatelescope.org/},
 GMRT\footnote{http://gmrt.ncra.tifr.res.in/}, OWFA\footnote{https://arxiv.org/abs/1703.00621}, MEERKAT\footnote{http://www.ska.ac.za/meerkat/},  MWA\footnote{https://www.mwatelescope.org/}, CHIME\footnote{http://chime.phas.ubc.ca/} aims to 
measure this weak signal \citep{chang2010hydrogen, masui2013measurement, switzer2013determination}.
At low redshifts $ z < 6$ following the complex epoch of reionization \citep{Gallerani_2006}, the \nh distribution is believed to be primarily housed in self-shielded DLA systems \citep{wolfe05, proch05}. The post reionization \nh 21-cm modeled as a tracer of the underlying dark matter distribution, quantified by a bias \citep{Bagla_2010, Guha_Sarkar_2012, Sarkar_2016, Carucci_2017} and a mean neutral fraction (which does not evolve with redshift) \citep{xhibar, xhibar1, xhibar2}. Several works report the possibility of extracting cosmological information from the post-reionization 21-cm signal \citep{poreion1, mcquinn2006cosmological, poreion0, Mao_2008, poreion2, poreion3, poreion4, poreion5, poreion6, poreion7, poreion8, poreion9, poreion10, poreion11, poreion12, poreion13}.

The possibility of 21-cm intensity mapping  experiments as a precision probe of cosmology faces several observational challenges. The signal is buried is foregrounds from galactic and extragalactic sources.
 While the foregrounds are  several orders of magnitude
brighter than the 21 cm signal, its spectral properties are strikingly different from the 21-cm signal which allows for the two to be separated.
Several techniques attempt to remove the
foregrounds from the measured visibilities (e.g. \citep{paciga2011gmrt, datta2010bright, chapman2012foreground, mertens2018statistical, trott2022multi}, by  assuming the smooth nature of the foregrounds.
The multi-frequency angular power spectrum (MAPS) \citep{datta2007multifrequency} has been proposed as a tool for foreground removal by several groups \citep{2011MNRAS.418.2584G, 2023MNRAS.525.3439E}.
Some other groups adopt a ‘foreground avoidance’ strategy where
only the region outside the foreground wedge is used to estimate the 21-cm powerspectrum (e.g.\cite{pober2013opening, pober2014next, liu2014epoch, dillon2015empirical, pal2021demonstrating})  
Further, one requires extremely precise  bandpass, calibration for detection of the signal. 
Calibration introduce spectral structure
into the foreground signal, making it further difficult to effectively remove foregrounds. 
This difficulty has led to many proposals for precise bandpass calibration \citep{mitchell2008real, kazemi2011radio, sullivan2012fast, kazemi2013robust, dillon2020redundant, kern2020absolute, byrne2021unified, sims2022bayesian, ewall2022precision, byrne2023delay}.

In this paper, we have made projections of uncertainties on the dark energy parameters in Quintessence models, with a  non zero vacuum, using a proposed future observation of the power spectrum of the post-reionization 21 cm signal.
We have used a Fisher/Monte-carlo analysis to indicate how the error projection on the binned power spectrum allow us to constrain dark energy models with a negative $\Lambda$.

The paper is organized as follows: In Section-2 we discuss the dark energy models and  constraints of  observable quantities like the Hubble parameter and growth rate of  density perturbations from diverse observations. In Section-3 we discuss the 21-cm signal from the post reionization epoch and noise projections using the futuristic SKA1 -mid observations. 
We also constrain  dark energy model parameters  using Markov Chain Monte Carlo (MCMC) simulation. We discuss our results and other pertinent observational issues in the concluding section.

\section{Quintessence dark energy with non-zero vacuum}
We consider a Universe where the Quintessence field ($\phi$) and cosmological constant $\Lambda$ both contribute to the overall dark energy density i.e. $\rho_{DE} = \rho_\phi + \Lambda$ with the constraint that $\rho_{DE} > 0$ to ensure the late time cosmic acceleration \citep{sen2023cosmological}.
 Instead of working with a specific form of the Quintessence potential we chose to use a broad equation of state (EoS) parametrization $w_{\phi}(z)$. It has been shown that at most a two-parameter model can be optimally constrained from
observations \citep{PhysRevD.72.043509}. We use the CPL model proposed by \citet{CHEVALLIER_2001} and \citet{PhysRevLett.90.091301} which gave a phenomenological model-free parametrization and  incorporate several features of dark energy. This model has been extensively used by the Dark Energy Task force \citep{albrecht2006report} as the standard two parameter description of dark energy dynamics. 
It has also been shown that a wide class of quintessence scalar field models can be mapped into the CPL parametrization \citep{PhysRevD.93.103503}.
The equation of state (EoS)  is given by \[ w_\phi (z) = w_0 + w_a \left ( \frac{1}{1+z} \right ).\]
This model gives a smooth variation of 
$w_\phi(z) = w_0 + w_a ~ {\rm as} ~ z\rightarrow \infty ~~~ {\rm to} ~w_\phi(z) = w_0 ~~{\rm for} ~ z = 0 $
and the corresponding density of the quintessence field varies with redshift as $ \rho_\phi (a) \propto a^{-3(1+w_0 + w_a)} \exp^{3w_a a}$. 
 In a spatially flat Universe,  evolution of the Hubble parameter $H(a)$ is given by
\begin{equation}
    \frac{H(a)}{H_0}   = \sqrt{\Omega_{m_0}a^{-3}+ \Omega_{\phi 0}  ~{\rm exp}\left[ -3\int_1^a da' \frac{1+w_\phi(a')}{a'}\right] + \Omega_{\Lambda 0} }
\end{equation}
with  $\Omega_{m0} + \Omega_{\phi 0} + \Omega_{\Lambda} = 1$. 
We shall henceforth call this model with $\Lambda$ along with a scalar field as  the CPL-$\Lambda$CDM model.

We consider two important cosmological observables. 
Firstly we consider  a dimensionless quantifier of cosmological distances \citep{Eisenstein_2005}
\be
r_{_{BAO}} (z) = \frac{r_s}{D_{_V}(z)} \ee
where $r_s$ denotes the sound horizon at the drag epoch and $D_{_V} (z)$ is the BAO effective distance  $D_V$  \citep{amendola_tsujikawa_2010} is defined as 
\be
D_{_V} (z) = \left [(1 + z)^2 D^2_A(z)  \frac{c z}{H(z)} \right ]^{1/3} \ee
This dimension-less distance $r_{_{BAO}} $ is a quantifier of the background cosmological model (density parameters) and is thereby sensitive to the dynamical evolution of dark energy.

Secondly we use the growth rate of density fluctuations as a quantifier of cosmological structure formation.
Clustering of galaxies in spectroscopic surveys \citep{zhao2021completed}, counts of galaxy clusters \citep{campanelli2012galaxy, sakr2022cluster} aim to measure the quantity called 
the growth rate of matter density perturbations  and the root mean square
normalization of the matter power spectrum $\sigma_8$ given by:
\begin{equation}
\label{eq:growth}
f (a) \equiv
\frac{d~\log~D_+ (a)}{d~\log~a}  ~~ \mathrm{and}~~\sigma_8(a) \equiv \sigma_{8,0} \frac{D_+ (a)}{D_+ (a=1)}
\end{equation} 
A more robust and reliable quantity $f\sigma_8 (a)$ that is measured by
redshift surveys is the combination of the growth rate
$f(a)$ and $\sigma_8(a)$. 
\begin{figure}
\begin{center}
\includegraphics[scale=0.335]{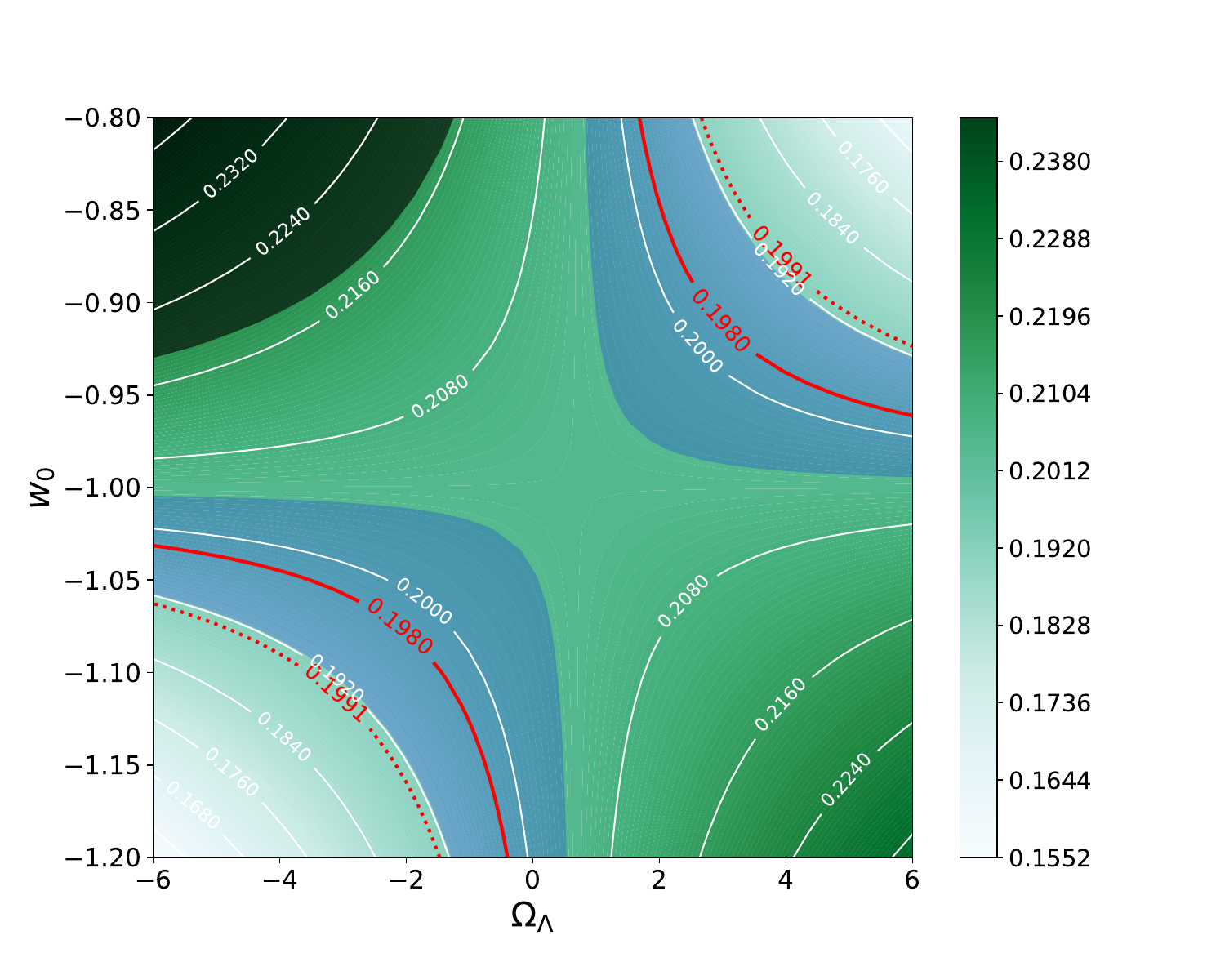}
\includegraphics[scale=0.335]{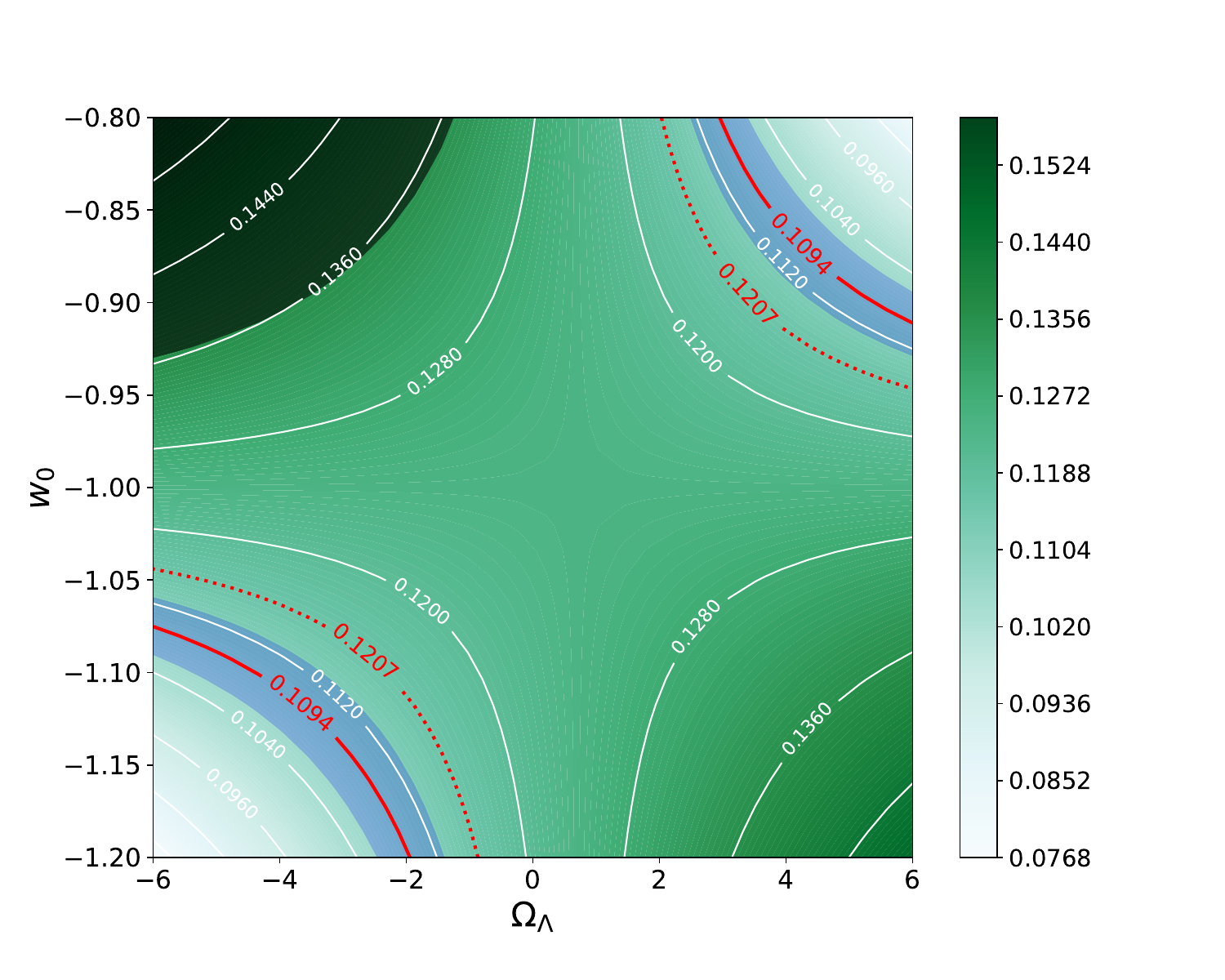}\\
\includegraphics[scale=0.315]{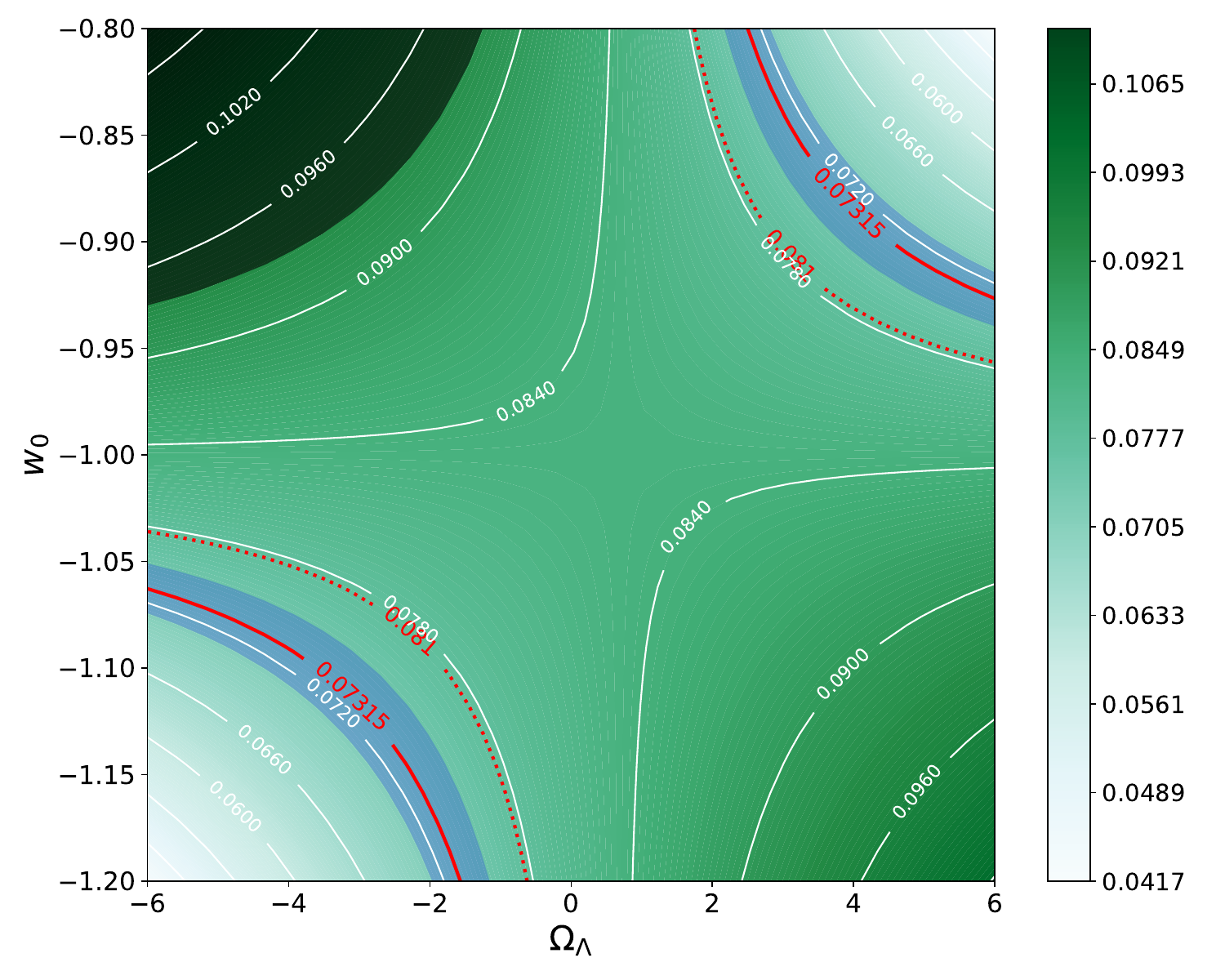}
\caption{shows $r_{BAO}$ in the $(\Omega_{\Lambda}, w_0)$ plane. The red contour line corresponds to the observational data point and the blue shaded region depicts the $1\sigma$ errors. The data points in the left two figures come from the 2df galaxy survey at redshifts of $z = 0.2$ and $ z= 0.35$ respectively \citep{Percival_2007} and the third figure shows the high redshift data  at $z = 0.57$ from BOSS SDSS-III survey \citep{Anderson_2012}. The red dotted contour  correspond to $r_{BAO}$ computed for a $\Lambda$CDM model. The grey sectors correspond to the models for which the Universe did not ever go through an accelerated phase till that redshift.}
\label{fig:fig1}
\end{center}
\end{figure}
\begin{figure}
\begin{center}
\includegraphics[scale=0.335]{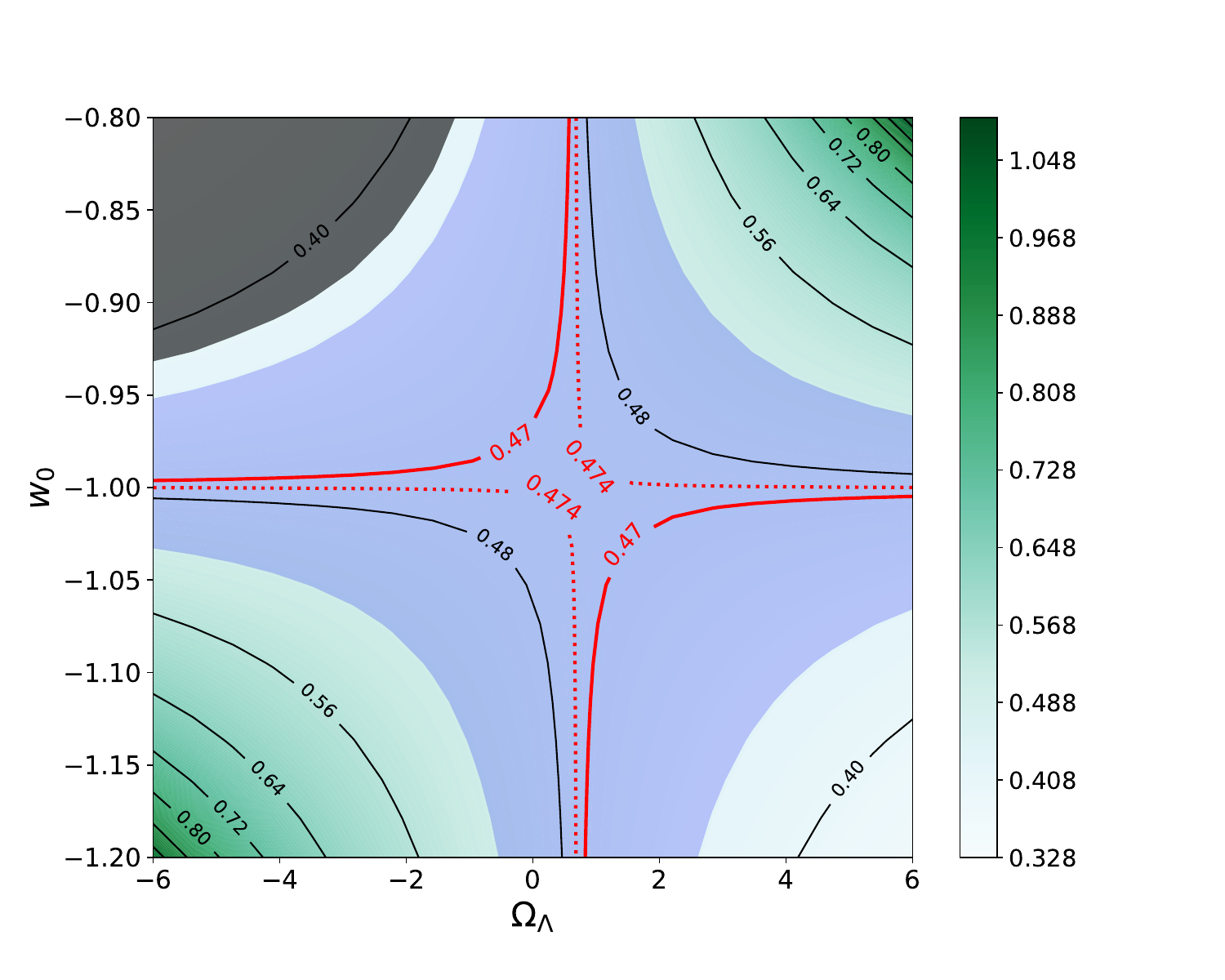}
\includegraphics[scale=0.335]{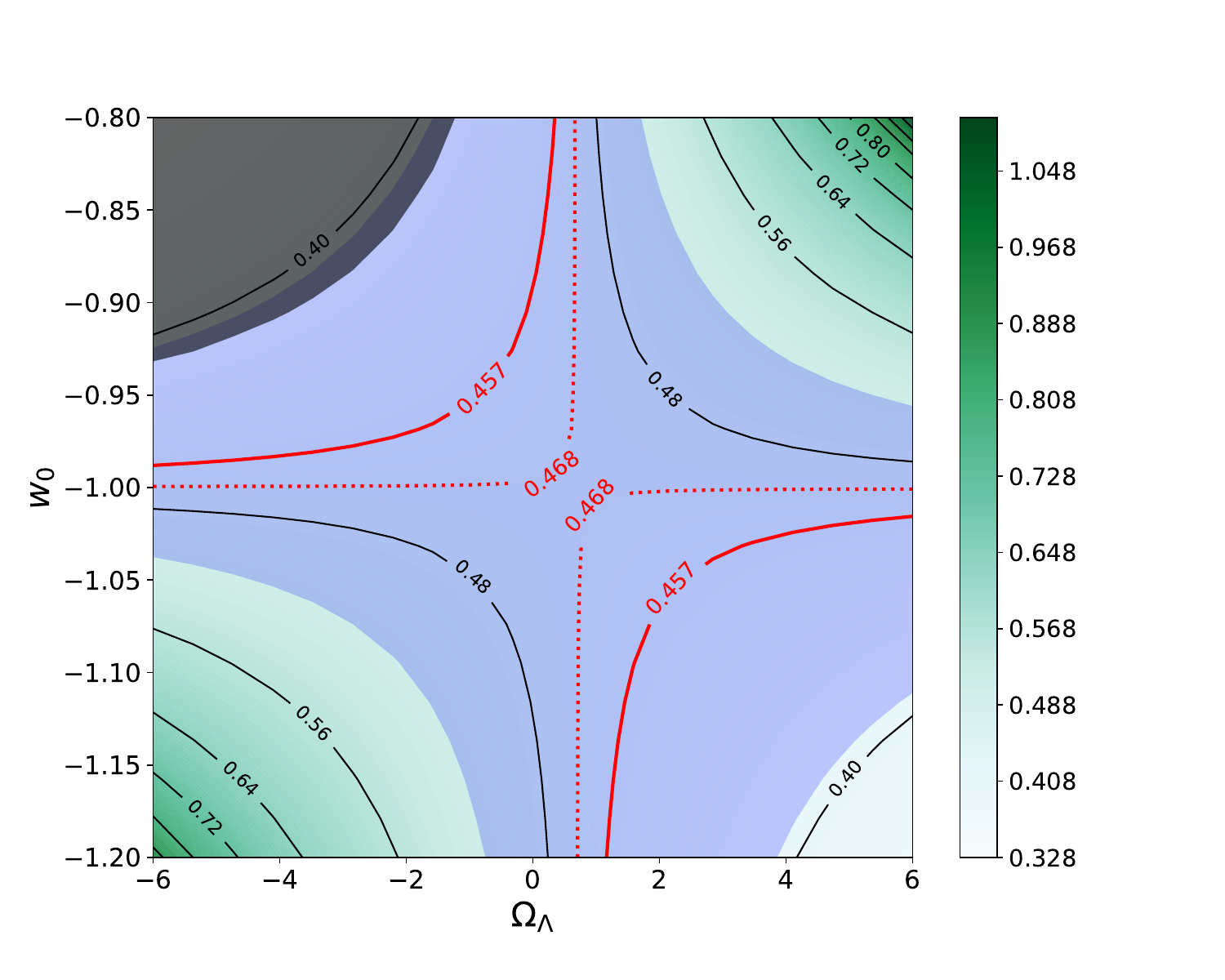}
\includegraphics[scale=0.315]{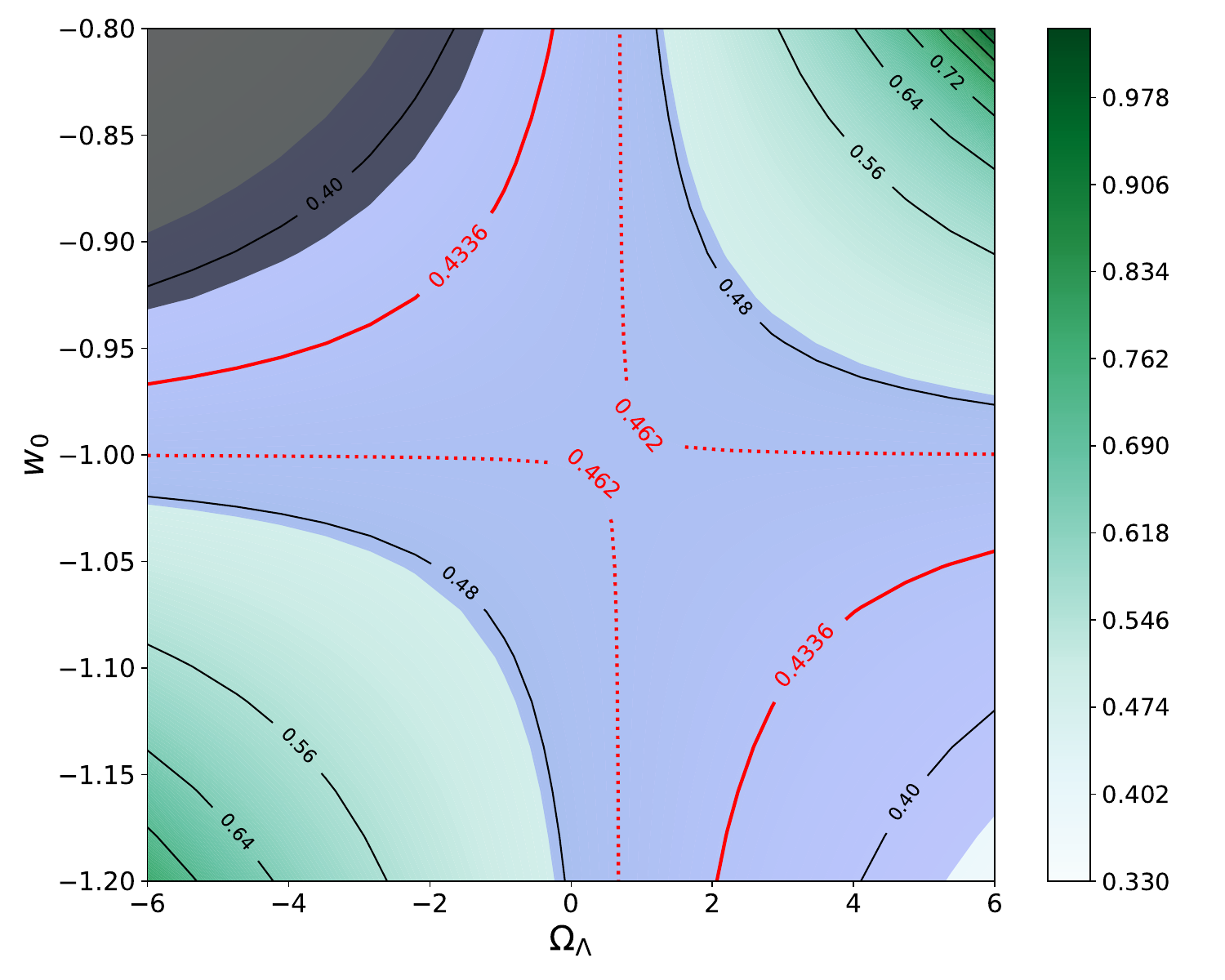}
\caption{shows variation of $f\sigma_8(z)$ in the $(\Omega_\Lambda , w_0)$ plane. The solid red line corresponds to the observational data points  from  SDSS-III BOSS $f\sigma_8(z=0.51) = 0.470 \pm 0.041$ \citep{sanchez2017clustering},  $f\sigma_8(z=0.61)= 0.457 \pm 0.052$ \citep{chuang2017clustering} and eBOSS DR16 LRGxELG data $f\sigma_8(z=0.7)= 0.4336 \pm 0.05003$ \citep{zhao2021completed}.
The red dotted contour  corresponds to $f\sigma_8(z)$ computed for a $\Lambda$CDM model.
 The grey sectors correspond to the models for which the Universe did not ever go through an accelerated phase till that redshift.
}
\label{fig:fig2}
\end{center}
\end{figure}
Figure (\ref{fig:fig1}) shows variation of $r_{BAO}$ in the $(\Omega_{\Lambda}, w_0)$ plane for the CPL-$\Lambda$CDM model with $H_0 = 72$ Km/s/Mpc. We have chosen $w_a =0$ for simplicity. 
 Further we have kept the $r_s$ fixed to the value computed for the fixed $\Omega_m$ and $\Omega_b$ from $\Lambda$CDM model \cite{Planck2018}. We note that  $r_s$ does not change much with $\Omega_{\phi}$ $\Omega_{\Lambda}$. 

We note that $\Omega_{\Lambda}$ is negative in the second and third quadrant. 
The red contour line corresponds to the observational data and the blue shaded region depicts the $1\sigma$ errors. The first figure in the panel corresponds to $z = 0.2$ and the red contour corresponds to observations from the 2df galaxy redshift survey gives the bounds on $r_{_{BAO}}$ as $r_{_{BAO}}( z = 0.2) = 0.1980 \pm 0.0058$ \citep{Percival_2007}. 
The second figure in the panel corresponds to $z=0.35$ with measured  $r_{_{BAO}}(z = 0.35) = 0.1094 \pm 0.0033$ \citep{Percival_2007}.
 The analysis of BOSS (SDSS III) CMASS sample along with Luminous red galaxy sample \citep{Anderson_2012}  from SDSS-II  gives  $r_{_{BAO}}( z = 0.57) = 0.07315 \pm 0.002$, as is shown in the third figure of the panel. We also show the contour for $r_{BAO}$ at the corresponding to that redshift for a pure $\Lambda$CDM cosmology with 
cosmological parameters \citep{Planck2018} results  $ (\Omega_{m{_0}},\Omega_{b_{0}},H_0,n_s,\sigma_8, \Omega_K) = (0.315,~0.0496,~67.4,~0.965,~0.811, ~0) $.
 All these observations are consistent with the possibility of models with negative $\Lambda$ with varying uncertainties. It is clear from the observations  that there are two separate
regions consistent with data: The third quadrant corresponds to Phantom models with negative $\Lambda$  and the first quadrant which corresponds to  non-phantom models with positive $\Lambda$. It is also clear that in spatially flat cosmologies with conditions $\rho_m > 0 $
and $\rho_{\phi} >0$  implies that $\Omega_{\Lambda} < 1 $ which is not supported by data.
The addition of a negative cosmological constant to a phantom dark energy model seems viable from the data. We find that the CPL-$\Lambda$ CDM  with a phantom field and negative $\Lambda$  and $H_0 = 72$ Km/s/Mpc,  the observational data as also  $\Lambda$CDM with $H_0 = 67.4$Km/s/Mpc are all qualitatively consistent.
We note that while computing $r_{BAO}$, the sound horizon distance  $r_s$ is fixed to the value computed for  $\Omega_m$ and $\Omega_b$ from \cite{Planck2018} since $r_s$ does not change much with $\Omega_{\phi}$ $\Omega_{\Lambda}$.

Figure (\ref{fig:fig2}) shows variation of $f\sigma_8(z)$ in the $(\Omega_\Lambda , w_0)$ plane. The solid red line corresponds to the observational data  from  SDSS-III BOSS $f\sigma_8(z=0.51) = 0.470 \pm 0.041$ \citep{sanchez2017clustering},  $f\sigma_8(z=0.61)= 0.457 \pm 0.052$ \citep{chuang2017clustering} and eBOSS DR16 LRGxELG data $f\sigma_8(z=0.7)= 0.4336 \pm 0.05003$ \citep{zhao2021completed} respectively.
While the mean observational $f\sigma_8$ data falls in the non-phantom sector with negative $\Lambda$, the error bars are quite large and again, 
the $\Lambda$CDM  predictions (with $H_0 = 67.4$Km/s/Mpc), observed data and  CPL-$\Lambda$CDM with phantom field and negative $\Lambda$ for $H_0 = 72$ Km/s/Mpc are all consistent within $1-\sigma$ errors. The addition of a  negative $\Lambda$ to a phantom dark energy model seems to also push $H_0$ to a higher value.

In models with negative cosmological constants there are regions in the $(w_0- \Omega_{\Lambda})$ which corresponds to cosmologies which never had an accelerated phase in the past or had a transient accelerated phase or $H^2(z) <0 $. These regions are studied in an earlier work \cite {calderon2021negative}. In the range of $(w_0- \Omega_{\Lambda})$ shown in the above figures we have shaded these  regions where the acceleration parameter became negative, corresponding to the fact that in these models, the universe did not ever accelerate upto that redshift.

\section{The Post-reionioztion \nh 21-cm signal }

The epoch  reionization epoch  is believed to have ended around $ z \sim 6$ \citep{Gallerani_2006}. Subsequently only a  small fraction of \nh survives the process of ionization and remains housed in the over-dense regions of the IGM. These neutral clumped, dense  gas clouds remain neutral and shielded from background ionizing radiation. These are now believed to be the damped Lyman-{$\alpha$} systems (DLAs) \citep{wolfe05} associated with galaxies. The predominant source of the 21-cm radiation in epochs $z <6$ are these DLA system which  stores   $\sim 80\%$ of the \nh  at $z<4$ \cite{proch05} with \nh column density greater than $ 2 \times 10^{20}$atoms/$\rm cm^2$ \citep{xhibar, xhibar1, xhibar2}. The study of  clustering of  DLAs  indicate  their association with galaxies. These gas clumps are hence have a biased presence  in regions where matter over densities are highly  non-linear \cite{coo06, zwaan, nagamine}. The possibility of the presently functioning and upcoming radio telescopes to detect the cosmological 21-cm signal from low redshifts has led to an extensive literature on the post-reionization \nh signal  \citep{poreion9, poreion6,  poreion1, poreion2, poreion7, poreion8, poreion0}. Though  flux from individual DLA clouds is extremely  weak ($< 10\mu \rm Jy$) to be detected in radio observations, even with the next generation radio arrays, it is possible to detect the collective diffuse radiation  without requiring to resolve the individual sources.
Such an intensity mapping of the  diffused background in all radio-observations at the observation
frequencies less than $1420$MHz is believed to give a wealth of cosmological and astrophysical information.
Measuring the statistical properties of the fluctuations of the diffuse 21-cm intensity distribution on the plane of the sky and as a function of redshift  gives a  way to study cosmological structure formation tomographically. Modeling the post-reionization \nh signal is based on  several simplifying assumptions which are  supported by extensive numerical simulations and astrophysical observations.

\begin{itemize}
\item {\it Post-reionization 21-cm Spin temperature : }

In the post-reionization epoch the spin temperature $T_s >> T_{\gamma}$ where $T_{\gamma}$ is the  CMB
temperature. This is due to the Wouthheusen field coupling which leads to an enhanced population of the triplet state of \nh. Consequently radiative transfer of CMBR through a gas cloud in this epoch shall cause the 21-cm
radiation is seen in emission against the background CMBR \citep{madau97, bharad04, zaldaloeb}.
Further,  the  kinetic gas temperature remains strongly coupled to the Spin temperature through Lyman-$\alpha$
scattering or collisional coupling \citep{madau97}.
\item {\it Mean neutral fraction:} 

Lyman-{$\alpha$} forest studies indicate  that the value of the density parameter of the neutral gas  is $ \Omega_{gas} \sim 10^{-3}$ for $ \approx 1 \leq z \leq 3.5$ \citep{proch05}.
Thus the  mean neutral
fraction  is $\bar{x}_{\rm HI} = \Omega_{gas}/\Omega_b \sim 2.45\times 10^{-2}$.
This value remains constant in the post-reionization epoch for $ z \leq 6$.

\item {\it Peculiar flow of \nh: }

The theory of cosmological perturbation shows that on  large sacles  the  baryonic matter falls into the regions of dark matter overdensities. Thus the non-Hubble  \nh peculiar flow of the gas is primarily  determined by the dark
matter distribution on large scales.
The \nh peculiar velocity manifests as a  redshift space
distortion  anisotropy in the 21-cm power spectrum in a manner similar to the Kaiser effect seen in galaxy surveys \citep{hamilton1998linear}.

\item { \it Intensity mapping and noise due to discrete clouds:}
The source of the 21-cm signal are DLA clouds. Intensity mapping ignores the discrete nature of the sources and aims to map the smoothed diffuse intensity distribution \citep{Furlanetto_2006, Pritchard_2012, Bull_2015}. The discreteness of the source  will introduce a Poisson sampling noise. We neglect this noise in our modeling since the number density $n$ of the DLA sources is very large \citep{poreion8} and the Poisson noise typically goes as $1/n$.

\item {\it Gaussian fluctuations:}

The overdensity field of dark matter distribution is believed to be  generated by Gaussian process in the very early Universe leading to  a scale invariant primordial power spectrum. We assume that there are no non-gaussianities, whereby the statistics of the random overdensity field is completely exhausted by studying the two-point correlation/power-spectrum. All $p$-point correlation functions where $p$ is odd, are assumed to be zero in the first approximation. Primordial non-gaussianity and non-linear structure formation will make the field non-gaussian, but this is neglected as a first approximation. The gas is believed to follow the dark matter and also expected to not show any  non-gaussian effects.

\item   { \it Post-reionization \nh as a biased tracer:}

The distribution of baryonic matter in the form of neutral hydrogen is an unsolved problem in cosmology. The linear theory predictions indicate that on large scales, baryonic matter follows the underlying dark matter distribution. However, at low redshifts, the growth of density fluctuations is likely to be plagued by non-linearilites and it is not {\it apriori} meaningful to extrapolate the predictions  
of linear theory in this epoch where overdensities $\delta \sim 1$.
Galaxy redshift surveys show that the galaxies trace the underlying dark matter
distribution \citep{dekel, mo, yosh} with a bias. If we model the  post-reionization \nh to be primarily stored in  dark matter haloes,  it is plausible to expect that the gas to trace the underlying dark matter density field  with a possible bias as well.

We define a bias function $b_T(k, z)$ as $$
   {b}_T(k, z) = {\left [\frac{P_{\rm HI}(k, z)}{P_m(k, z)}\right ]}^{1/2}
$$ 
where  ${P}_{\rm HI}(k, z)$ and ${P}_m(k, z)$ denote the \nh and
dark matter power spectra respectively. 
With this definition of a general function  $b_T(k, z)$, we merely relocate the lack of knowledge of \nh distribution to a scale and redshift dependent function that quantifies the properties  of post-reionization \nh clustering.

 Theoretical considerations show that the bias is scale dependent on small scales below  the Jean's length \citep{fang}. However, on large scales the bias is expected to be
scale-independent.
The scales above which the linear bias approximation is acceptable  is however, dependent on the redshift. 
While the neutral fraction on the post-reionization epoch is believed to be a constant,  studies \citep{poreion0} show that small  fluctuations in the ionizing background
may also contribute to a scale dependency in the bias  $b_T(k,z)$.
The most compelling studies of the post-reionization \nh has been through the use of N-body numerical simulations \citep{Bagla_2010, Guha_Sarkar_2012, Sarkar_2016, Carucci_2017}.
These  simulations uses diverse rules for  populating  neutral hydrogen to dark matter halos in a certain mass range and identifying them as DLAs. 

Similar to the behaviour of galaxy bias \citep{fry, dekel, mo, moo}, these N-body simulations of the post-reionization \nh agree on the generic qualitative behaviour.
On  large scale the bias is found to be linear (scale independent)  and is  a monotonically rising function of redshift for $1< z< 4$ \citep{Mar_n_2010}. 
However, on small scales the bias becomes scale-dependent  as rises steeply on small scales. The rise of the bias on small scales owes it origin to the  absence of small mass halos as is expected from the CDM power spectrum and consequent distribution of \nh in larger mass halos. In this work we use the fitting formula for $b_T(k, z)$  obtained from numerical simulations \citep{Sarkar_2016}. 

\end{itemize}

\subsection{The post-reionization \nh 21cm power spectrum }
Adopting all the modeling assumptions discussed in the last section,  the power spectrum of post-reionization  \nh  21-cm  excess brightness temperature field $\delta T_b$ from redshift
$z$ \citep{Furlanetto_2006, Bull_2015, bharad04, param3} is given by
\begin{equation}
P_{21}(k, z, \mu)  = \mathcal{A}^2_T {( b_T  + f \mu^2)}^2  P_m(k, z)
\end{equation}
where 
\begin{equation} 
\A_{T} = 4.0 \, {\rm {mK}} \,
b_{T} \, {\bar{x}_{\rm HI}}(1 + z)^2\left ( \frac{\Omega_{b0}
  h^2}{0.02} \right ) \left ( \frac{0.7}{h} \right) \left(
\frac{H_0}{H(z)} \right) 
\end{equation}
The term $ f(z)  \mu^2$ has its origin in the  \nh peculiar
velocities \citep{poreion2, bharad04} which, is also assumed to be sourced
by the dark matter fluctuations.

Since our cosmological model is significantly different from the fiducial one (i.e., $\Lambda$CDM), 
the difference will introduce additional anisotropies in the correlation function through the Alcock-Paczynski effect \citep{simpson2010difficulties,samushia2012interpreting,montanari2012new}.
In the presence of the Alcock-Paczynski effect, the redshift-space HI 21-cm power spectrum is given by: \citep{Furlanetto_2006, Bull_2015}
\begin{equation}
\label{21cmps}
P_{21}(k, z, \mu) = \frac{\A^2_T}{\alpha_\parallel \alpha^2_\perp} \left[ b_T + \frac{f(z)\mu^2}{F^2 + \mu^2 (1-F^2)}   \right]^2  P_m \left( \frac{k}{\alpha_\perp} \sqrt{1+\mu^2 (F^{-2}-1)}, z \right)
\end{equation}
where $F = \alpha_\parallel / \alpha_\perp$, with $\alpha_\parallel$ and $\alpha_\perp$ being the ratios of angular and radial distances between fiducial and real cosmologies, $\alpha_\parallel = H^f/H^r$, $\alpha_\perp =D_A^r/D_A^f$. 

The overall factor $\alpha_\parallel \alpha^2_\perp$ is due to the scaling of the survey's physical volume.  As the real geometry of the Universe differs from the one predicted by the  fiducial cosmology, we introduce additional distortion in the redshift space. 
The AP test is sensitive to the isotropy of the Universe and can help differentiate between different cosmological models.
We note that the geometric factors shall also imprint in the BAO feature of the power spectrum.
Since $0 \leq \mu  1 $ the redshift space 21cm power spectrum can be decomposed in the basis of Legendre polynomials $\mathscr{P}_\ell (\mu)$ as \citep{hamilton1998linear}
\begin{equation}
P_{21}(k,\mu , z) = \sum_\ell P_{\ell}(z , k) \mathscr{P}_\ell (\mu)
\label{eq:exp}
\end{equation}
The odd harmonics vanish by pair exchange symmetry and non-zero azimuthal harmonics.
( as $Y_{\ell , m}$'s with $m\neq 0$  vanish by symmetry about the line of
sight). Using the standard normalization 
\[
\int_{-1}^{+1}  \mathscr{P}_\ell (\mu)  \mathscr{P}_r (\mu) d \mu  = \frac{2}{2\ell + 1} \delta_{\ell, r} \]
the first few Legendre polynomials are given by 
\begin{equation}
\mathscr{P}_0 (\mu) = 1 , ~~ \mathscr{P}_2 (\mu) = \frac{1}{2} \left( 3 \mu^2 -1 \right) , ~~ \mathscr{P}_4 (\mu) = \frac{1}{8}(35\mu^4 -30\mu^2 + 3)
\end{equation}
The  coefficients of the expansion of the 21cm power spectrum, can be found by inverting the  equation (\ref{eq:exp}).  Thus we have 
\begin{equation}
P_\ell (z,k) = \frac{(2\ell + 1)}{2} \int_{-1}^{+1} d\mu ~ \mathscr{P}_\ell (\mu) P_{21} (z,k,\mu)
\end{equation}
While full information is contained in an infinite set of  functions $ \{ P_\ell (z,k) \} $, we shall be interested in the first few of these function which has the dominant information.

\subsection{The BAO feature in the multipoles of 21-cm power spectrum}
The sound horizon at the drag epoch  provides a standard ruler, which can be used to calibrate cosmological distances.  Baryons imprint the cosmological power spectrum through a distinct oscillatory
signature \citep{White_2005, Hu-eisen}.  The BAO imprint on the 21-cm
signal has been studied extensively \citep{sarkar2013predictions,sarkar2011imprint}. The baryon acoustic oscillation (BAO) is an important probe of cosmology
\citep{Eisenstein_2005, Percival_2007, Anderson_2012,
  shoji2009extracting, sarkar2013predictions} as it allows us to
measure the angular diameter distance $D_A(z)$ and the Hubble
parameter $H(z)$ using the transverse and the longitudinal
oscillatory features respectively \citep{lopez2014alcock}.

 The sound horizon at the drag epoch is given by 
\be
s(z_{d}) = \int_{0}^{a_{drag} } \frac{c_s da}{a^2 H(a)} 
\ee 
where $a_{drag} $ is the scale factor at the drag epoch redshift $z_d$ and
$c_s$ is the sound speed given by $c_s(a) = c/
\sqrt{3(1+3\rho_b/4\rho_\gamma)}$  where $\rho_b$ and $\rho_\gamma$
denotes the baryonic and photon densities, respectively. The Planck 2018 constrains the value of $z_{d}$ and $s(z_d)$ to be $z_{d}
= 1060.01 \pm 0.29$ and $s (z_{d}) = 147.21 \pm 0.23 $Mpc
\citep{Planck2018}. We shall use these as the fiducial values in our
subsequent analysis.  The standard ruler \lq$s$' defines a transverse
angular scale and a redshift interval in the radial direction as \be
\theta_s (z) =\frac{s(z_d)} {(1+z) D_A (z)} ~~~~~~~ \delta z_s =
\frac{s (z_d) H(z)}{c} \ee Measurement of $\theta_s$ and $\delta z_s$,
allows the independent determination of $D_A (z)$ and $H(z)$. The BAO feature comes from the baryonic part of $P(k)$. 
In order to isolate the BAO feature, we subtract the cold dark matter power spectrum from total $P(k)$ as $P_b (k) = P(k) - P_c(k)$. Owing to significant deviations between the assumed cosmology and
the fiducial cosmology, our longitudinal and tangential coordinates are rescaled by  $\alpha_{\parallel}$ and $\alpha_\perp$ respectively, the true power spectrum scaled as $k' = k\sqrt{1+\mu^2 (F^{-2}-1)}/\alpha_\perp$ from the apparent one \citep{matsubara1996, ballinger1996, simpson2010difficulties}. Incorporating the Alcock-Paczynski corrections explicitly in the BAO power spectrum can be written as
\citep{hu1996small, seo2007improved} \be
\label{eq:baops}
P_b (k') = A \frac{\sin x}{x} e^{-(k'\sum_s)^{1.4}}e^{-k'^2
  \sum_{nl}^2/2} \ee where $A$ is a normalization, $\sum_s =
1/k_{silk}$ and $\sum_s = 1/k_{nl}$ denotes the inverse scale of \lq
Silk-damping' and \lq non-linearity' respectively. In our analysis we
have used $k_{nl} = (3.07 h^{-1}Mpc)^{-1} $and $k_{silk} = (8.38
h^{-1}Mpc)^{-1} $ from \citet{seo2007improved} and $x =
\sqrt{k^2(1-\mu^2) s_\perp^2 + k^2\mu^2 s_\parallel^2}$. The changes in $D_A(z)$ and $H(z)$
are reflected as changes in the values of $s_\perp$ and $s_\parallel$
respectively, and the errors in $s_\perp$ and $s_\parallel$
corresponds to fractional errors in $D_A$ and $H(z)$ respectively. We
use $p_1 = \ln (s^{-1}_{\perp})$ and $p_2 = \ln (s_{\parallel})$ as
parameters in our analysis. The Fisher matrix is given by
\be 
F_{ij} = \left(\frac{2\ell +1}{2}\right) \int dk' ~ \int_{-1}^{+1} d\mu ~ \frac{A^2_T}{\alpha_\parallel \alpha^2_\perp} \left[ b_T + \frac{f(z)\mu^2}{F^2 + \mu^2 (1-F^2)} \right]^2 \frac{\mathscr{P}_\ell(\mu)}{\delta P^2_{21}(k, z, \mu)} \frac{\partial P_b(k')}{\partial p_i}\frac{\partial P_b (k')}{\partial p_j} \ee
\be 
=  \left(\frac{2\ell +1}{2}\right) \int dk' ~ \int_{-1}^{+1} d\mu ~  \frac{A^2_T}{\alpha_\parallel \alpha^2_\perp} \left[ b_T + \frac{f(z)\mu^2}{F^2 + \mu^2 (1-F^2)} \right]^2 \frac{\mathscr{P}_\ell(\mu)}{\delta P^2_{21}(k, z, \mu)}  \left( \cos x - \frac{\sin x}{x}\right)^2 f_i f_j A^2 e^{-2(k'\sum_s)^{1.4}}e^{-k'^2
  \sum_{nl}^2} 
\ee
where $f_1 = \mu^2 -1$ and $f_2 = \mu^2 $.

We choose SKA's a Medium-Deep Band-2 survey that covers a sky area of 5,000 $\deg^2$
in the frequency range $0.95-1.75$GHz ($z =[0-0.5]$) and a Wide Band-1 survey that covers a sky area of 20,000 $\deg^2$ in the frequency range $0.35-1.05$GHz ($z = [0.35-3]$) \citep{bacon2020cosmology}. We calculate the expected error projections on $D_A(z)$ and $H(z)$ in five evenly spaced, non-overlapping redshift bins, in
the redshift range [z=0-3] with $\Delta z = 0.5$. Each of the six bins is taken to be independent and is centered at redshifts of $z = [0.25, 0.75, 1.25, 1.75, 2.25]$.

\subsection{Visibility correlation}
We use a visibility correlation approach to estimate the noise power spectrum for the 21-cm signal \citep{poreion1,bali,mcquinn2006cosmological,geil2011polarized,villaescusa2014modeling,Sarkar_2015}.
A radiointerferometric observation measures the complex visibility. The measured visibility written as a function of baseline $\u = (u, v) $ and frequency $\nu$ is  a sum of signal and noise
\be \v(\u, \nu) = \S(\u, \nu ) + \N(\u, \nu)
\ee
\be \S(\u, \nu ) =  \frac{2k_B}{\lambda^2}  \int d \t ~A(\t) e^{2 \pi i \u \cdot \t} ~\delta T_b( \t, \nu)    \ee
where, $\delta T_b( \t, \nu)$ is the fluctuations of the 21-cm brightness temperature and $A(\t)$ is the telescope beam. The factor $ \left( \frac{2k_B}{\lambda^2}  \right )^2 $ converts brightness temperature to intensity (Raleigh Jeans limit).
Defining $\Delta \nu$ as the difference from the central frequency, a  further Fourier transform in frequency  $\Delta \nu$ gives us
\be s(\u, \tau ) =  \frac{2k_B}{\lambda^2}   \int d \t ~d \nu  ~A(\t) B( \Delta \nu) ~e^{2 \pi i (\u \cdot \t + \tau \Delta \nu) } ~\delta T_b( \t, \nu)    \ee
where $B(\Delta \nu)$ is the frequency response function of the radio telescope.
\be s(\u_a, \tau_m ) =  \frac{2k_B}{\lambda^2}  \int d \t ~d \Delta \nu \int \frac{d^3 \k }{(2 \pi)^3} ~e^{ -i (\kperp  r \cdot \t  + \kpar r' \Delta \nu ) }~A(\t) B( \Delta \nu) ~e^{2 \pi i (\u_a \cdot \t + \tau_m \Delta \nu) } ~\widetilde{\delta T_b}( \kperp, \kpar)   \ee
where the tilde denotes a fourier transform and $r' = dr(\nu)/d\nu$.
 
\be s(\u_a, \tau_m ) =  \frac{2k_B}{\lambda^2}  \int d \t ~d \Delta \nu \int \frac{d^3 \k }{(2 \pi)^3} ~e^{ -i (\kperp r - 2 \pi \u_a ) \cdot \t }  e^{ -i (\kpar r' - 2 \pi \tau_m )\Delta \nu}  ~A(\t) B( \Delta \nu)  ~\widetilde{\delta T_b}( \kperp, \kpar)   \ee
Performing the $\t$ and $\Delta \nu $ integral we have

\be s(\u_a, \tau_m ) = \frac{2k_B}{\lambda^2}    \int \frac{d^3 \k }{(2 \pi)^3}  ~\widetilde{A}   \left (\frac{\kperp  r}{2 \pi}  -  \u_a \right ) \widetilde{B} \left ( \frac{\kpar r'}{2 \pi} -  \tau_m \right )   ~\widetilde{\delta T_b}( \kperp, \kpar)   \ee


Defining new integration variables as  $ \u =\frac{\kperp  r}{2 \pi}$ and  $ \tau = \frac{\kpar r'}{2 \pi} $ we have 
\begin{eqnarray}
\langle s(\u_a, \tau_m )    s^*(\u_b, \tau_n ) \rangle = \left ( \frac{2k_B}{\lambda^2}  \right )^2 \frac{1}{r^2 r'} \int d \u  ~d \tau    \widetilde{A}   \left (\u-  \u_a \right ) \widetilde{A} ^*  \left (\u   -  \u_b \right )  \widetilde{B} \left ( \tau  -  \tau_m \right )  \widetilde{B}^* \left ( \tau -  \tau_n  \right )   P_{21} \left ( \frac{2 \pi \u }{r}, \frac{2 \pi \tau }{r'} \right ) 
\end{eqnarray} 

Approximately, we may write
\be \int \widetilde{B} \left ( \tau  -  \tau_m \right )  \widetilde{B}^* \left ( \tau -  \tau_n \right ) \approx B \delta_{m,n} ~~~~\mathrm{and} ~~~~ \int d \u  ~    \widetilde{A}   \left (\u-  \u_a \right ) \widetilde{A} ^*  \left (\u   -  \u_b \right )   \approx \frac{\lambda^2}{A_e} \delta_{a,b} \ee
where $B$ is the bandwidth of the telescope
and
where $A_e$ is the effective area of each dish. Hence
\begin{equation}
\langle s(\u_a, \tau_m )    s^*(\u_b, \tau_n ) \rangle \approx  \left ( \frac{2k_B}{\lambda^2}  \right )^2 \frac{ B \lambda^2 }{r^2 r' A_e} ~~     P_{21} \left ( \frac{2 \pi \u_a }{r}, \frac{2 \pi \tau }{r'} \right ) \delta_{m,n}  \delta_{a, b} 
\end{equation} 
The noise in the visibilities measured at different baselines and frequency
channels are uncorrelated. We then have

\be
\langle \N(\u_a, \nu_m )~ \N^*(\u_b, \nu_n ) \rangle = \delta_{a, b} \delta_{m,n} 2 \sigma^2 \ee
where 
 \be
 \sigma = \frac{\sqrt 2 k_B T_{sys} }{A_e \sqrt{\Delta \nu  t }} \ee
 where $A_e$ is the effective area of the dishes, $t$ is the correlator integration time and $\Delta \nu$ is the channel width.
 If $B$ is the observing bandwidth, there would be $B/\Delta \nu$ channels. The system temperature $T_{sys}$ can be written as 
 \be T_{sys} = T_{inst} + T_{sky} \ee
 where
 \be T_{sky} = 60 {\rm K}  \left ( \frac{\nu }{300 ~{\rm MHz}} \right ) ^{-2.5} \ee
 
 Under a Fourier transform 
 \be
 n (\u, \tau  ) = \sum_{i = 1}^{B/\Delta \nu}     \N(\u, \nu_i ) \Delta \nu ~~ e^{2 \pi i \nu_i \tau } \ee
 

\be  \langle n ( \u_a,  \tau) ~n^*({\u_b, \tau}) \rangle= 2 \sigma^2   \delta_{a, b} \Delta \nu ^2  \frac{B}{\Delta \nu}    =   2 \sigma^2   \delta_{a, b} \Delta \nu  B \ee

\be  \langle n ( \u_a,  \tau) ~n^*({\u_b, \tau}) \rangle=
\frac {4   k_B^2 T_{sys}^2 B  }{A_e^2 t }  =   \left (   \frac { 2 k_B}{\lambda^2} \right ) ^2  \left (  \frac{ \lambda ^2 T_{sys} }{ A_e} \right ) ^2 \frac{B}{t}   \ee
Now cosidering a total observation time $T_o$  and a bin $\Delta \u$,  
there is a reduction of noise by a factor $\sqrt N_p$ where $N_p$ is  the number of visibility pairs in the bin 
\be
N_p = N_{vis} ( N_{vis} - 1 ) /2  \approx  N_{vis} ^2 /2 \ee
where $N_{vis}$ is the number of visibilities in the bin. We may write 
\be N_{vis} = \frac{N_{ant} ( N_{ant} - 1) } {2} \frac{T_o}{t} \rho(\u ) \delta^2 U  \ee
where $N_{ant} $ is the total number of antennas and $\rho(\u)$ is the baseline distribution function.
\be  \langle n ( \u_a,  \tau) ~n^*({\u_b, \tau}) \rangle=
  \left (   \frac { 2 k_B}{\lambda^2} \right ) ^2  \left (  \frac{ \lambda ^2 T_{sys} B}{ A_e} \right ) ^2   \frac{2 \delta_{a,b}  }{N_{ant} ( N_{ant} - 1 )  B ~T_o~ \rho(\u) \delta ^2U }\ee
where an additional reduction by $\sqrt 2$ is incorporated by considering visibilities in half plane.
The 21 cm power spectrum is not spherically symmetric, due to redshift space distortion but is symmetric around the polar angle $\phi$. Because of
this symmetry, we want to sum all the Fourier cells in
an annulus of constant $(k, ~ \mu = \cos \theta = \kpar / k )$  with radial width $\Delta k$  and
angular width $\Delta \theta$  for a statistical detection. The number
of independent cells in such an annulus is
\be 
N_c  = 2 \pi k^2 \sin ( \theta) \Delta k \Delta \theta \frac{Vol}{(2\pi)^3} = 2 \pi k^2  \Delta k \Delta \mu \frac{Vol}{(2\pi)^3}  \ee
where
\be
Vol = \frac{r^2 \lambda^2 r' B }{ A_e} \ee
Thus the full covariance matrix for visibility correlation is \citep{villaescusa2014modeling, Sarkar_2015, geil2011polarized,mcquinn2006cosmological}
\be 
C_{a,b} = \frac{1}{\sqrt N_c}  \left ( \frac{2k_B}{\lambda^2}  \right )^2 \left [  \frac{ B \lambda^2 }{r^2 r' A_e} ~~     P_{21} \left ( \frac{2 \pi \u_a }{r}, \frac{2 \pi \tau }{r'} \right )  +    \left (  \frac{ \lambda ^2 T_{sys} B}{ A_e} \right ) ^2   \frac{2  }{N_{ant} ( N_{ant} - 1 )  B ~T_o~ \rho(\u) \delta ^2U } \right ] \delta_{a, b}
\label{eq:noise}
\ee
We choose $ \delta ^2U = A_e/ \lambda^2$, $ \Delta k  = k/10$, $\Delta \mu = \mu/ 10$. 

The baseline distribution function $\rho(\u)$ is normalized as
\be
\int d\u \rho(\u) = 1 \ee
For uniform baseline distribution 
\be \rho (\u) = \frac{1}{\pi ( U_{max}^2  - U_{min}^2) } \ee
Generally 
\be \rho (\u) = c \int d^2 {{\bf r}} \rho_{ant} ({\bf r}) \rho_{ant} ({\bf r} - \lambda \u ) \ee
Where $c$ is fixed by normalization of $\rho(\u)$ and $\rho_{ant}$ is the  distribution of antennae. 
The covariance matrix in Eq (\ref{eq:noise}) is used in our analysis to make noise projections on the 21-cm power spectrum and its multipoles.
Observations with total time time exceeding a limiting value will make the instrumental noise insignificant and the  Signal to Noise Ratio is primarily influenced by cosmic variance for such observations. Therefore, by introducing $N_{point}$ as the number of independent pointings, the covariance is further reduced by a factor of $1/\sqrt{N_{point}}$.

\section{Results and discussion}

\begin{figure}
\begin{center}
\includegraphics[scale=0.28]{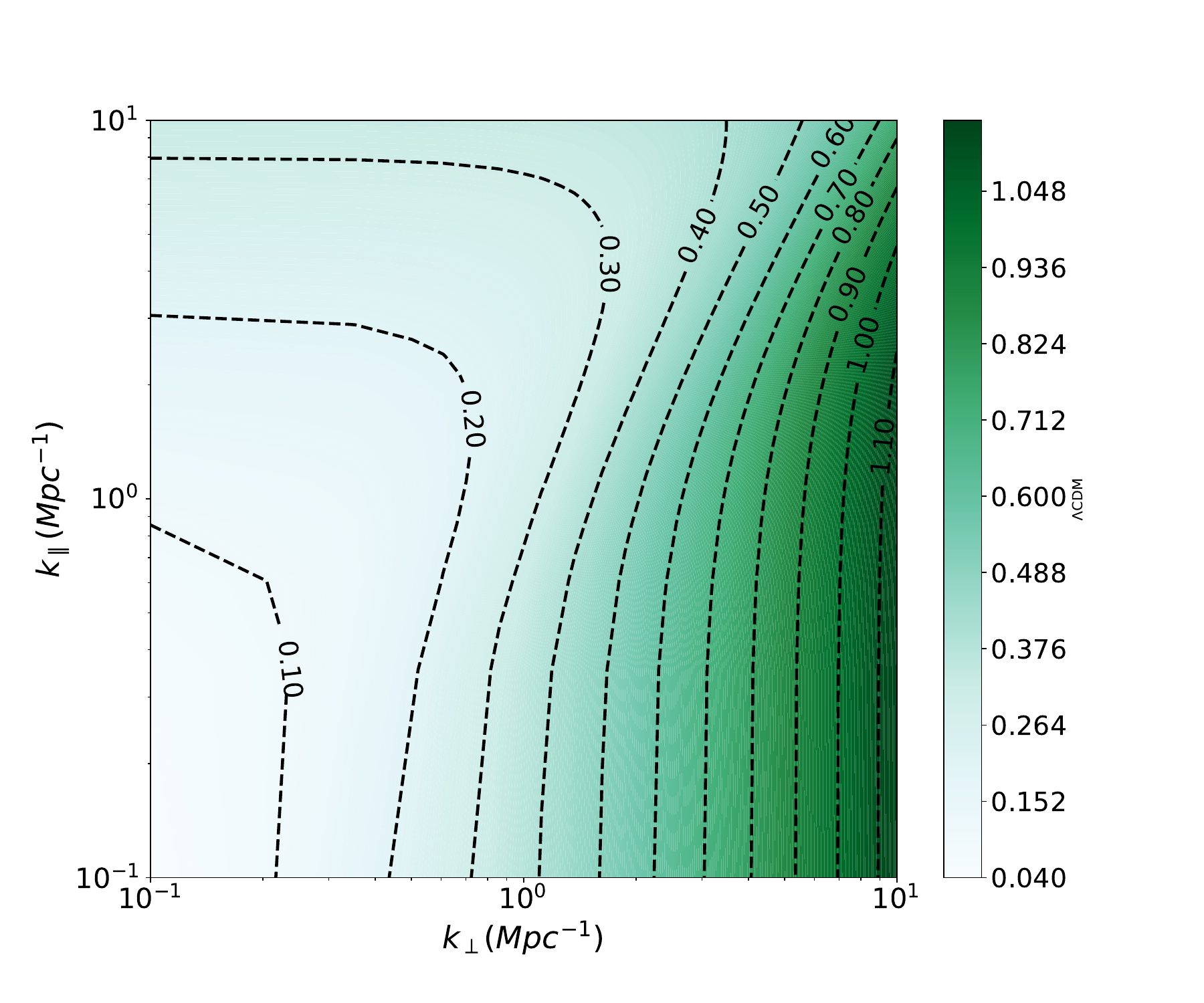}
\includegraphics[scale=0.28]{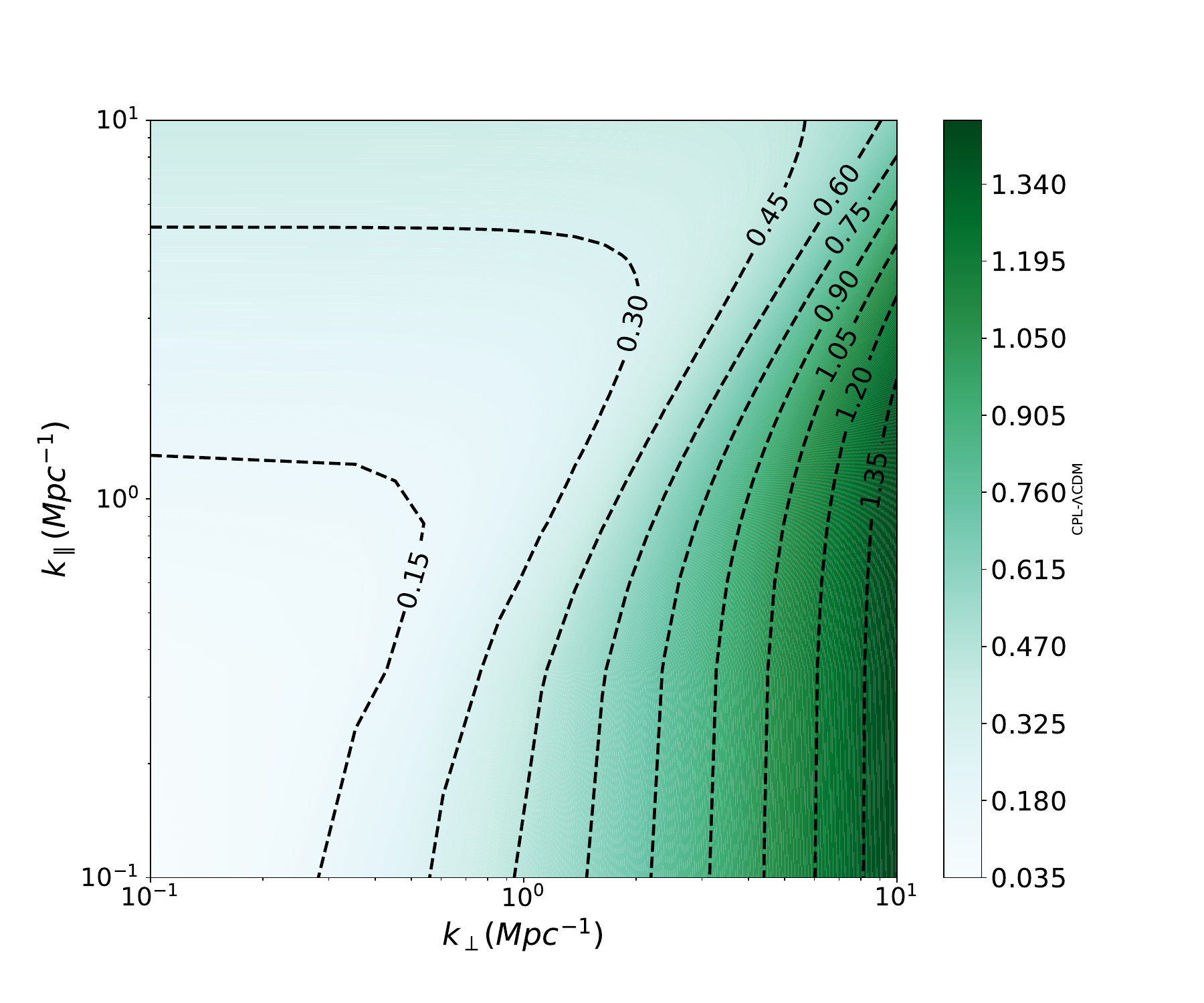}
\caption{ shows the 3D \nh 21-cm power spectrum at $z = 1$ in the ($k_\perp, k_\parallel$) space. The asymmetry in the signal is indicative of redshift space distortion: the left figure  corresponds to the $\Lambda$CDM. In contrast, the  right figure represents the CPL-$\Lambda$CDM model, where the Alcock-Paczynski effect enhanced the distortions. The colorbar shows the value of the dimensionless quantity $\Delta_{21}^2 = k^3P_{21}({\bf{k}})/(2\pi^2)$ in mK$^2$.}

\label{fig:3dps}
\end{center}
\end{figure}

\begin{table} [h]
\begin{minipage}{.8\linewidth}
\centering
\begin{tabular}{c |c | c | c | c | c }
\hline \hline
 $N_{ant}$  & Antennae Efficiency & $D_{dis}$ & $T_{o}$ & $T_{sys}$ & $B$  \\ [0.5ex] 
 \hline
 $250$ & 0.7 & $15$m & $500$hrs & $60$K & $200$MHz \\
 \hline
 \hline
  \end{tabular}
  \captionsetup{justification=centering}
\caption{Table showing the telescope parameters used in our analysis.}
\label{tab:noise-params}
\end{minipage}
\end{table}

\begin{figure}
\begin{center}
\includegraphics[scale=0.312]{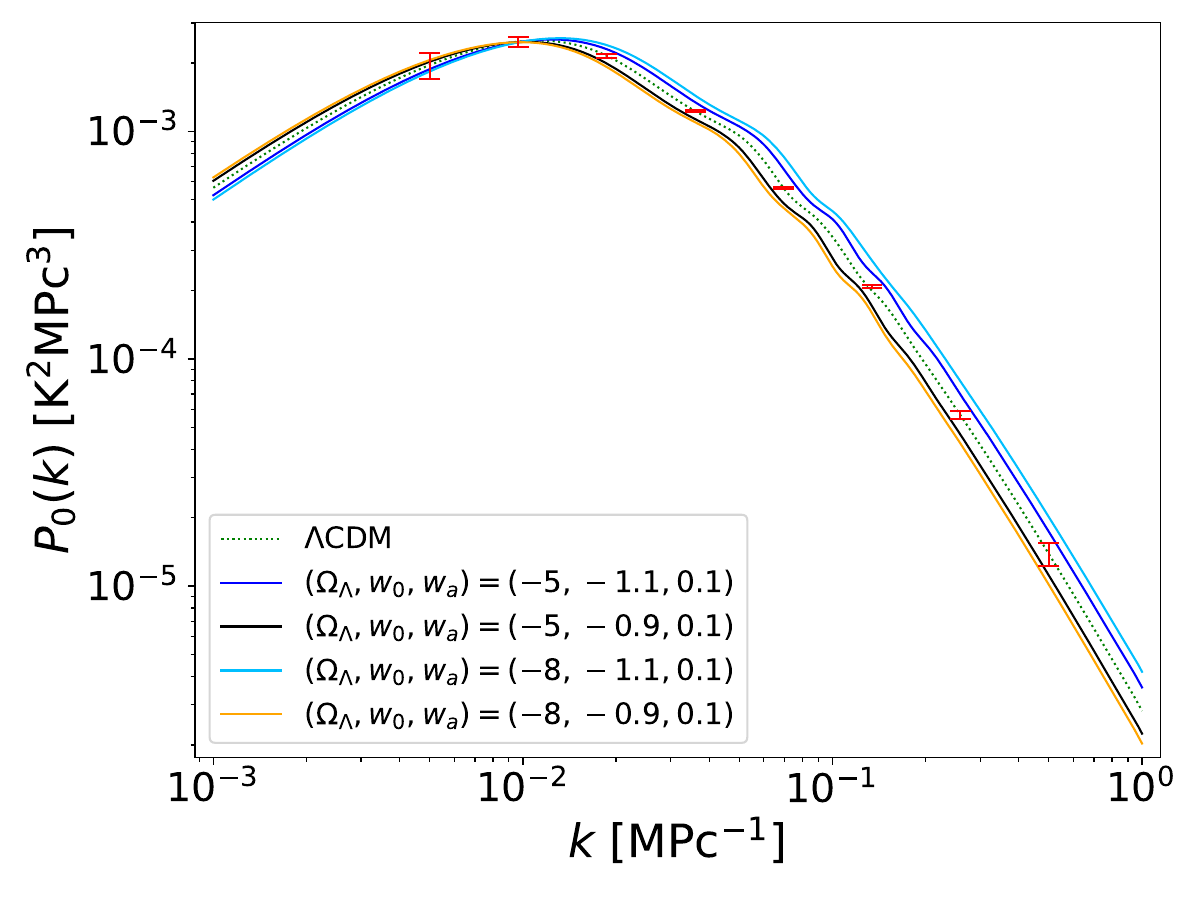}
\includegraphics[scale=0.312]{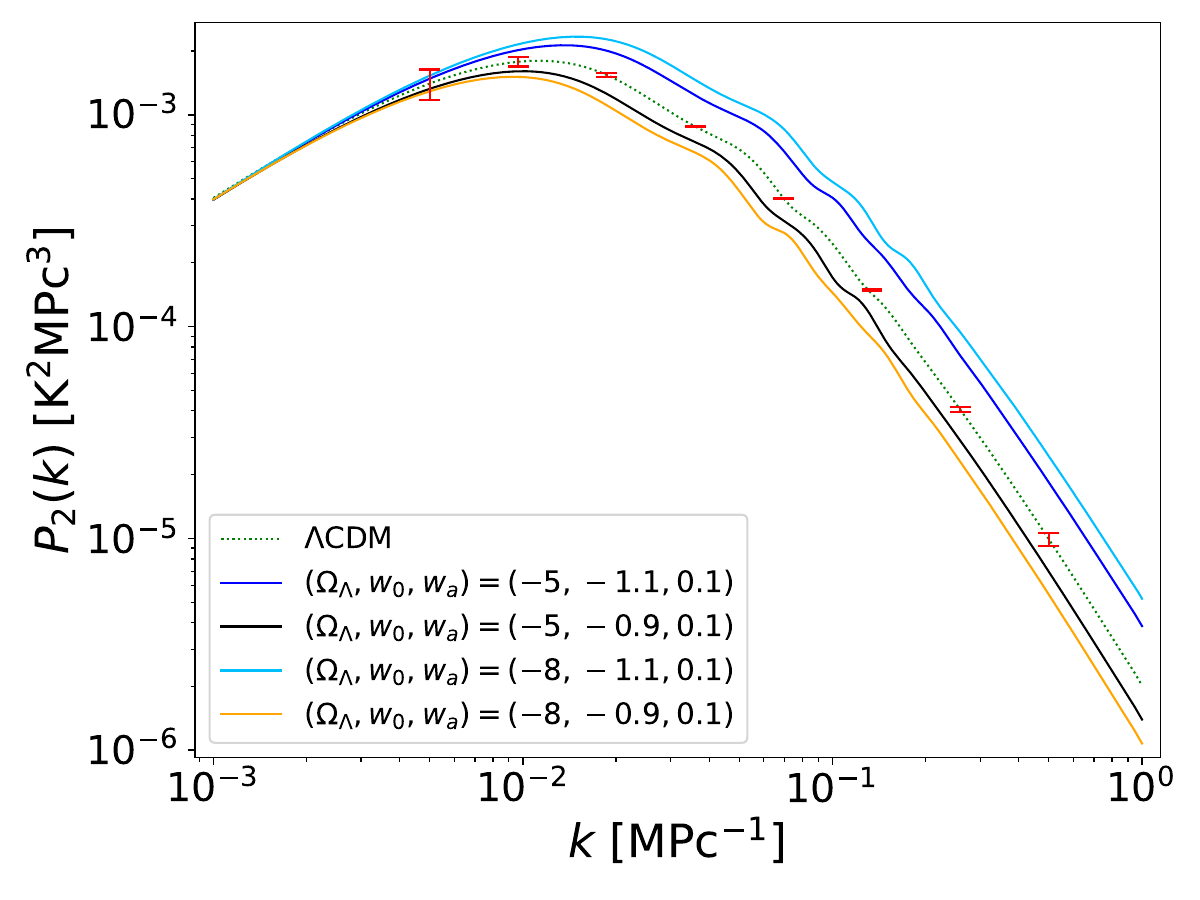}\\
\includegraphics[scale=0.312]{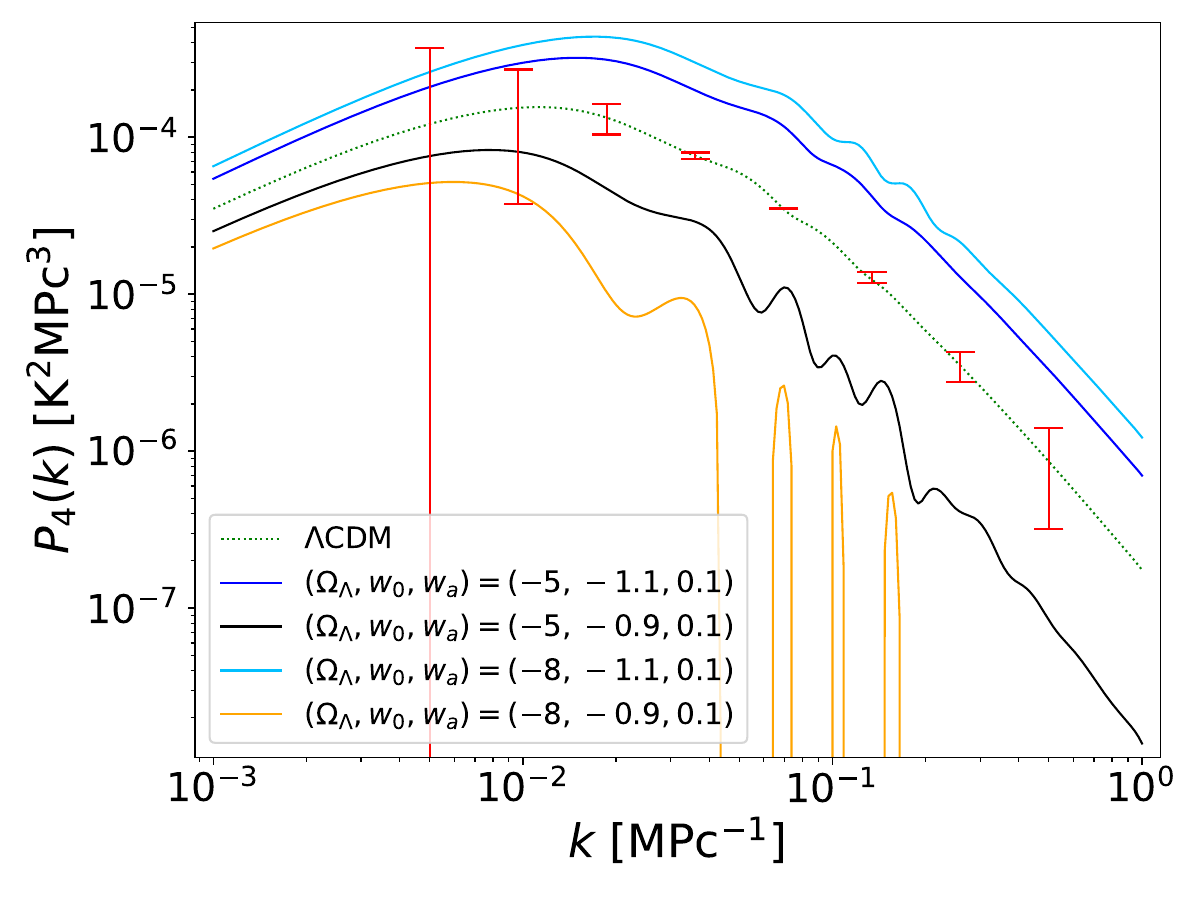}\\
\caption{shows the 21-cm linear power spectrum monopole (top-left), quadrupole (top-right) and hexadecapole (bottom) at redshift $z = 0.2$. The dotted line corresponds to $\Lambda$CDM.}
\includegraphics[scale=0.312]{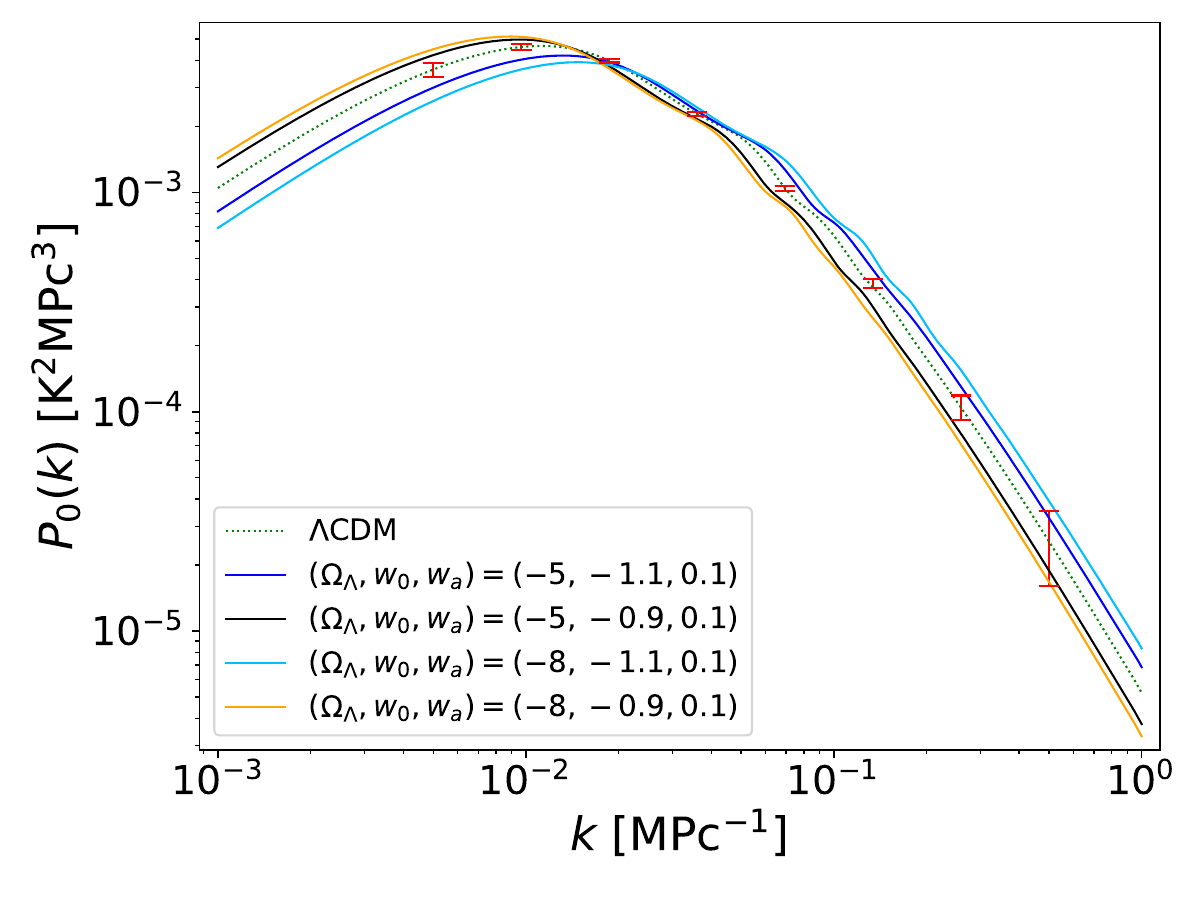}
\includegraphics[scale=0.312]{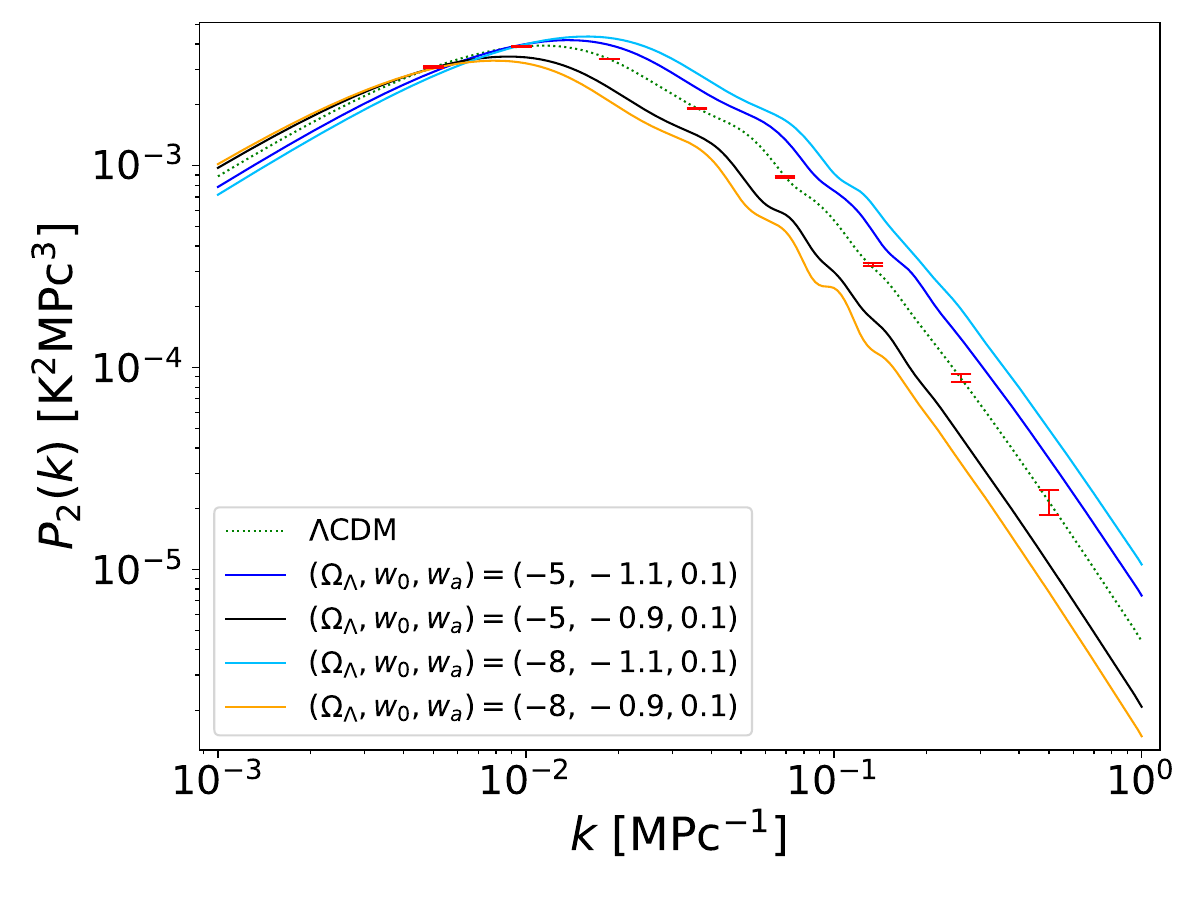}\\
\includegraphics[scale=0.312]{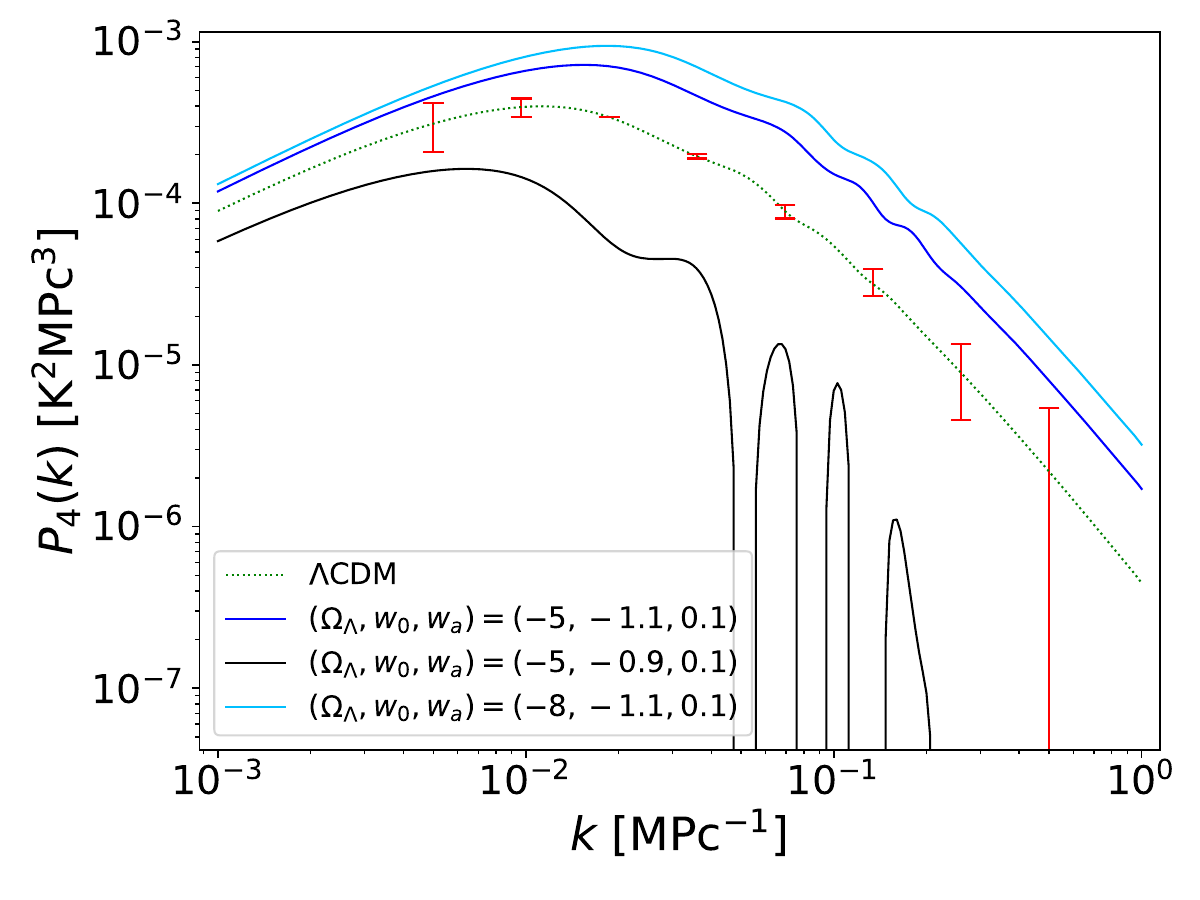}
\caption{shows the 21-cm linear power spectrum monopole (top-left), quadrupole (top-right) and hexadecapole (bottom) at redshift $z=0.57$. The dotted line corresponds to $\Lambda$CDM.}
\label{fig:multipole-signal}
\end{center}
\end{figure}
\begin{figure}
\begin{center}
\includegraphics[scale=0.5]{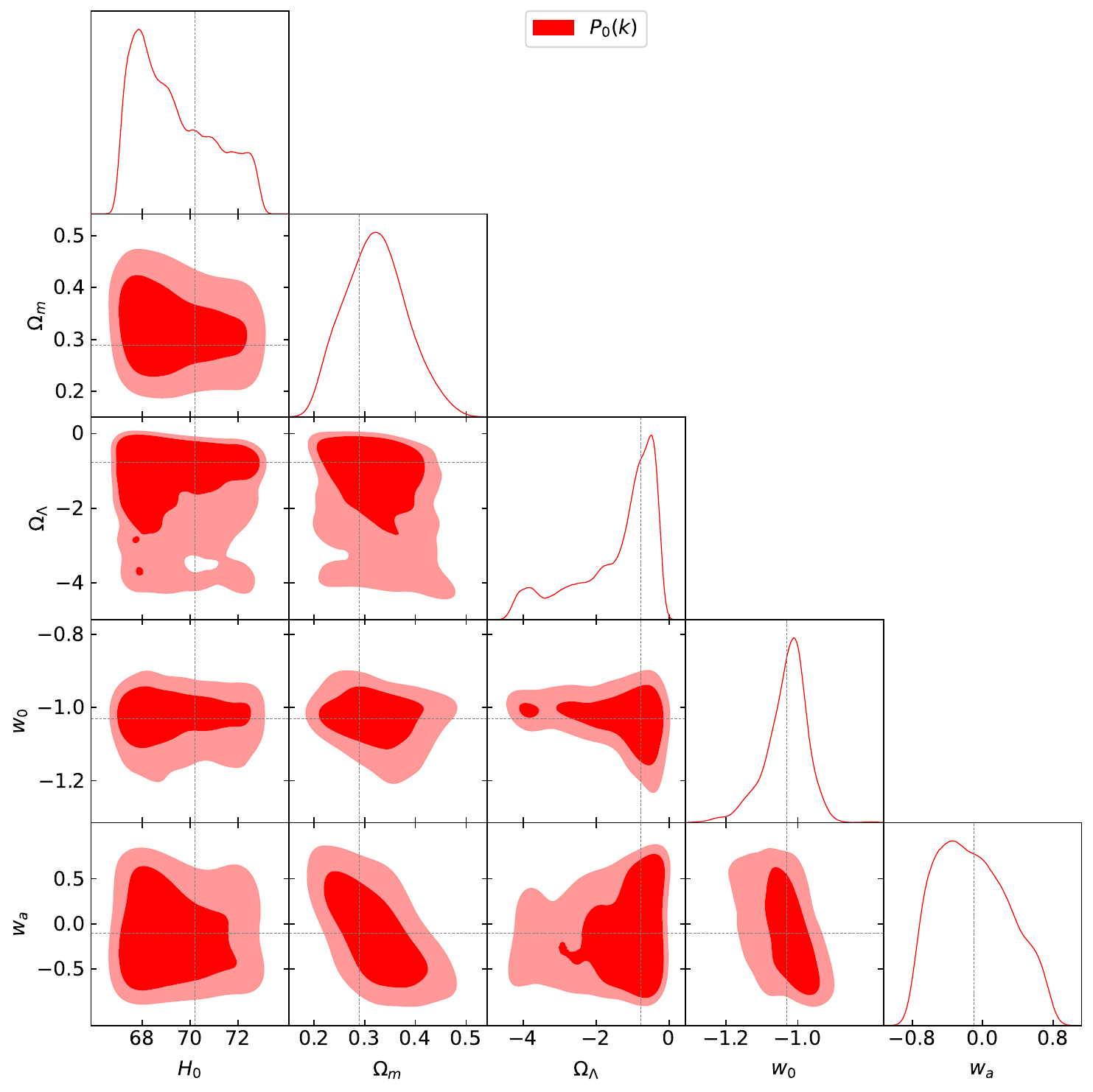}
\caption{  Marginalized posterior distribution of the set of parameters and ($H_0, \Omega_m, \Omega_\Lambda ,  w_0, w_a$) corresponding 2D confidence contours obtained from the MCMC analysis. The fiducial model parameters are taken from \citet{sen2023cosmological}} 
\label{fig:mcmc}
\end{center}
\end{figure}
In this section we discuss the results of our investigation. 
The figure (\ref{fig:3dps}) shows the dimensionless 3D 21-cm power spectrum $(\Delta_{21}^2 = k^3P_{21}({{\bf{k}},z})/(2\pi^2))$ in redshift space at the fiducial redshift $z = 1$. In the plane of $k_\parallel$ and $k_\perp$, the power spectrum shows the anisotropy of the redshift space power spectrum. The contours colored in blue correspond to the fiducial $\Lambda$CDM model, while those in red pertain to the CPL-$\Lambda$CDM model. We choose the best-fit value on CPL-$\Lambda$CDM model parameters ($\Omega_{m} = 0.289, \Omega_{\Lambda} = -0.781, w_0 = -1.03, w_a = -0.10$) obtained from the combined data CMB+BAO+Pantheon+R21  \citep{sen2023cosmological}. The Alcock-Paczynski effect makes a notable contribution, intensifying the anisotropy observed in the power spectrum. 
The significant departure of the CPL-$\Lambda$CDM model $\sim 5\%$ at $k \sim 1 Mpc^{-1}$ indicates that a closer investigation of the possibility of discerning such models from the $\Lambda$CDM model is justified.

For the measurement of the 21-cm power spectrum we consider a radio-interferometric observation using a futuristic  SKA1-Mid like experiment. 
The typical telescope parameters used are summarized in the table below. We also assume that the antenna distribution falls off as $1/r^2$, whereby the baseline coverage on small scales is suppressed.

We consider $250$ dish antennae
each of diameter $15$m and efficiency $0.7$. We assume $T_{sys} = 60K$ and an observation bandwidth of $128$MHz. The $k$-range between the smallest and largest baselines is binned as $\Delta k = \alpha k$ where $\alpha = 1/N_{bin} \ln (U_{max}/ U_{min})$. The minimum value of $k$ is taken to be $0.005$Mpc$^{-1}$ the maximum value of $k$ is taken to be $0.5$Mpc$^{-1}$ with logarithmically number of bins $N_{bin} = 8$. We consider a total observation time of $500\times 150$hrs with $150$ independent pointings, we obtain the $1-\sigma$ errors on $P_\ell (k,z)$. The fiducial model is chosen to be the $\Lambda$CDM. Figure (\ref{fig:multipole-signal}) shows the multiples of $P_{21}(k, z)$ for selective parameter values of CPL-$\Lambda$CDM model. The central dotted line corresponds to $\Lambda$CDM. The fiducial redshift is chosen to be $0.2$ (top) and $0.57$ (bottom). We found that in the $k$ range $0.01$Mpc$^{-1}<k<0.1$Mpc$^{-1}$ phantom models are distinguishable from $\Lambda$CDM at a sensitivity of $>3\sigma$. For higher multipoles, they are even more differentiable from fiducial $\Lambda$CDM. On the contrary, non-phantom models remain statistically indistinguishable from the $\Lambda$CDM model while considering monopole only. They are only distinguishable in higher multipoles. 

We see a strong  effect of $\Omega_\Lambda$ on the multipole components of the power spectrum. A non-trivial $\Omega_\Lambda$ introduces additional enhancement of anisotropy in the 21-cm power spectrum through the redshift space distortion factor $f \mu^2$. Additionally the power spectrum gets further modified  through the departure of the  factor $F = \alpha_{\parallel}/\alpha_{\perp}$ from  unity and through the matter power spectrum $P(k,\mu)$ though the scalings of $k_{\perp}$ and $k_{\parallel} $. This explains the significant deviation of the 21-cm power spectrum for the CPL-$\Lambda$CDM model from its  standard $\Lambda$CDM counterpart. This is become more prominent  in the  quadrupole and hexadecapole components cause of the terms with the anisotropy are enhanced by integrals of higher powers of  $\mu$ in  the Legendre polynomials. 

\begin{table}

\centering

\begin{tabular}{p{0.10\linewidth}p{0.10\linewidth}p{0.10\linewidth}p{0.10\linewidth}p{0.10\linewidth}p{0.10\linewidth}p{0.10\linewidth}}

\hline \hline

 Parameters &~~~~~~~~$H_0$~~~~~~ & $\Omega_{m}$ &  $\Omega_{\Lambda}$ & $~~~~~~~w_0$ & $~~~~~~~w_a$ \\ [0.5ex] 

 \hline\hline

Constraints  &~~ $68.957^{2.342}_{-1.306}$~~~~~~ & $0.320^{0.063}_{-0.064}$ & $-1.030^{0.589}_{-1.712} $ & $-1.023^{0.043}_{-0.060}$ & $-0.141^{0.478}_{-0.409}$ \\ 

 \hline

 \hline

 \hline

\end{tabular}

\caption{ The parameter values, obtained in the MCMC analysis are tabulated along the $1-\sigma$ uncertainty.}

\label{tab:MCMC-constraints}

\end{table}

The BAO imprint on the monopole $P_0(z,k)$ allows us to constrain $D_A(z)$ and $H(z)$. 
We perform a Markov Chain Monte Carlo (MCMC) analysis to constrain the model parameters using the projected error constraints obtained on the binned $H(z)$ and $D_A(z)$ from the $P_0(z, k)$. The analysis uses the Python implementation of the MCMC sampler introduced by \citet{foreman2013emcee}. We take flat priors for CPL-$\Lambda$CDM model parameters with ranges of $H_0 \in [67, 73], \Omega_m \in [0.2,0.6], \Omega_{\Lambda} \in [-7, 2], w_0 \in [-1.5, 1.5], w_a \in [-0.7, 0.7]$. The figure (\ref{fig:mcmc}) shows the marginalized posterior distribution of the set of parameters ($H_0, \Omega_m , \Omega_\Lambda, w_0, w_a$), and the corresponding 2D confidence contours are obtained. The fiducial value of the model parameters are taken from the best fit values of $H_0 , \Omega_m , \Omega_{\phi}, \Omega_\Lambda , w_0 , w_a$ obtained from the combined data CMB+BAO+Pantheon+R21 \citep{sen2023cosmological}. Constraints on model parameters are tabulated in Table (\ref{tab:MCMC-constraints}). While comparing with the projected error limits for the parameters of the CPL-$\Lambda$CDM as obtained in \citet{sen2023cosmological}, we find that 21-cm alone  doesn't impose stringent constraints on the values of $\Omega_\Lambda$ and $w_a$. However, it does exhibit a reasonably good ability to constrain the parameter $w_0$. To attain more robust constraints on these model parameters, a more comprehensive approach is required. This involves combining the 21-cm power spectrum data with other cosmological observations such as the CMB, BAO, SNIa, galaxy surveys etc. Through the joint analysis, it becomes possible to significantly improve the precision of parameter estimation.

\section{Conclusion}
 In this work, we study the possibility of constraining  negative $\Lambda$ using the post-reionization  \nh 21-cm power spectrum. We specifically investigate the quintessence models with the most widely used dark energy EoS parameterization and add a non-zero vacua (in terms of a $\pm  \Lambda$). 

By the analysis of BOSS (SDSS) data we find that addition of a negative cosmological constant to a phantom dark energy model seems viable.
We see that the CPL-$\Lambda$CDM  with a phantom field and negative $\Lambda$  and $H_0 = 72$ Km/s/Mpc qualitatively consistent with the data.

Further we study the non-trivial  CPL-$\Lambda$CDM  model with the $f\sigma_8$ data from the galaxy surveys. We find that the mean observational $f\sigma_8$  falls in the non-phantom sector with negative $\Lambda$. 
Since the error bars are quite large, both  $\Lambda$CDM  predictions (with $H_0 = 67.4$Km/s/Mpc), and  CPL-$\Lambda$CDM with phantom field and negative $\Lambda$ for $H_0 = 72$ Km/s/Mpc are consistent within $1-\sigma$ errors. The addition of a  negative $\Lambda$ to a phantom dark energy model also seems to push $H_0$ to a higher value.

Subsequently, we look into the influence of the Alcock-Packzynski effect on 3D \nh 21-cm power spectrum. Using $\Lambda$CDM as a fiducial cosmology, we explore the implications of the first few multipoles of the redshift-space 21-cm power spectrum for the upcoming SKA intensity mapping experiments. 
We find that the multipoles specially the quadrupole and hexadecapole components show significant departure from their standard $\Lambda$CDM counterparts. We focus on the BAO feature on the monopole component, and estimate the projected errors on the $H(z)$ and $D_A(z)$ over a redshift range $z \sim 0-3$. 

Further, we perform a MCMC analysis to constrain the CPL-$\Lambda$CDM model parameters using the projected error constraints obtained on the binned $H(z)$ and $D_A(z)$ from the $P_0(z, k)$. We find that 21-cm alone  doesn't impose stringent constraints on the model parameters. Combining the 21-cm power spectrum data with other cosmological observations such as the CMB, BAO, SNIa, galaxy surveys etc  can significantly improve the precision of parameter estimation.

We have not factored in several observational challenges towards detecting the 21-cm signal. Proper mitigation of large galactic and extra-galactic foregrounds and minimizing calibration errors are imperative for the any cosmological investigation. In a largely observationally idealized scenario, we have obtained error projections on the model parameters from the BAO imprint on the post-reionization 21-cm  intensity maps. We  employ a  Bayesian analysis techniques to put constraints on the model parameters. Precision measurement of these parameters shall enhance our understanding of the underlying cosmological dynamics and potential implications of negative $\Lambda$ values. 

\section*{Acknowledgements}

 AAS acknowledges the funding from SERB, Govt of India under the research grant no: CRG/2020/004347. 

\section{Data Availability}
The data are available upon reasonable request from the corresponding author.

\bibliographystyle{mn2e}
\bibliography{references}
\end{document}